\documentclass[trackchanges,twocolumn]{aastex701}
\usepackage{lipsum}
\hypersetup{linkcolor=red,citecolor=blue,filecolor=cyan,urlcolor=blue}
\graphicspath{{./}{Figures/}}

\received{\today}
\revised{\today}
\accepted{\today}
\submitjournal{ApJ}

\begin{document}

\title{Host--Non-host Differences in Stellar Chemistry, Activity, and Birth Radius: Hints of Distinct Formation Environments for Earth-like Planets and Sub-Neptunes}

\author[orcid=0000-0003-3957-9067]{Xunzhou Chen}
\affiliation{School of Science, Hangzhou Dianzi University, Hangzhou, PR China}
\affiliation{National Astronomical Data Center Zhijiang Branch, Hangzhou, PR China}
\email[show]{cxz@hdu.edu.cn}

\author[orcid=0000-0003-0795-4854]{Tiancheng Sun}
\affiliation{CAS Key Laboratory of Optical Astronomy, National Astronomical Observatories, Chinese Academy of Sciences, Beijing 100101, China}
\email[show]{suntc@bao.ac.cn}

\author[0000-0003-4769-3273]{Yuxi (Lucy) Lu}
\affiliation{Department of Astronomy, The Ohio State University, Columbus, 140 W 18th Ave, OH 43210, USA}
\affiliation{Center for Cosmology and Astroparticle Physics (CCAPP), The Ohio State University, 191 W. Woodruff Ave., Columbus, OH 43210, USA}
\email{lucylulu12311@gmail.com}

\author{Liu Long}
\affiliation{Department of Physics, Xiangtan University, Xiangtan 411105, Hunan Province, China}
\email{longliu@xtu.edu.cn}

\author{Zixuan Lu}
\affiliation{School of Physics and Astronomy, Beijing Normal University, Beijing, China}
\email{202231160008@mail.bnu.edu.cn}

\begin{abstract}

Host-star properties provide important clues to planet formation and evolution, yet it remains unclear whether the observed differences between planet-hosting stars and stars without detected planets reflect genuine signatures of planet formation or underlying differences in stellar populations. Using a homogeneous sample of 28,383 Kepler–LAMOST–Gaia dwarf and subgiant stars, including 629 host stars with 865 planets, we compare host stars with age–mass matched non-host stars after correcting for distribution-induced matching biases. Most host–non-host differences disappear when the full planet sample is considered. However, separating planets by radius reveals distinct trends. At the high-abundance end of the [O/Fe], [Mg/Fe], and [Si/Fe] distributions, Earth-like hosts tend to be more O-rich but more Mg- and Si-poor than their age–mass matched non-host stars, whereas sub-Neptune hosts tend to show the opposite behavior. Sub-Neptune hosts also tend to exhibit lower chromospheric activity and smaller birth radii than comparable non-host stars. These results suggest that host–non-host differences become apparent primarily after separating planets by size and that Earth-like planets and sub-Neptunes may be associated with distinct formation environments and evolutionary pathways. We also find that hot-Jupiter hosts are tentatively more metal-rich, chromospherically active, and formed at smaller Galactic birth radii than hosts of longer-period Jupiters.

\end{abstract}



\section{Introduction}
Over the past three decades, exoplanet discoveries have revealed that planetary systems are both common and remarkably diverse. Ground-based surveys and dedicated space missions, most notably \textit{Kepler}, have revolutionized the study of exoplanets by shifting the field from individual detections to population-level investigations based on statistically powerful samples \citep[e.g.,][]{2010Sci...327..977B,2011arXiv1109.2497M}. More than 6000 exoplanets have now been confirmed (NASA Exoplanet Archive; \citealt{2013PASP..125..989A}), spanning a broad range of planetary radii, masses, orbital periods, and system architectures. This rapidly growing census has enabled increasingly detailed investigations of the physical processes that govern planet formation and evolution.

A major approach to understanding planet formation and evolution has been to investigate how planetary systems depend on the properties of their host stars. Stellar chemistry is expected to influence planet formation by regulating the amount of solid material available in protoplanetary disks. Consistent with this picture, spectroscopic surveys have established a strong correlation between stellar metallicity and planet occurrence, particularly for giant planets \citep[e.g.,][]{2012MNRAS.423..122B,2013ApJ...771..107E,2014Natur.509..593B,2014ApJ...789L...3D,2015AJ....149..143F,2018AJ....155...89P,2021AJ....161..114S,2022AJ....164...60S}. This trend is commonly interpreted as evidence that metal-rich disks provide more favorable conditions for planet formation \citep[e.g.,][]{2004A&A...415.1153S,Johnson_2010}. Stellar age provides a temporal perspective on planetary-system evolution \citep[e.g.,][]{2021AJ....161..114S,David_2021,2023AJ....166..243Y,2025NatAs...9..995T,2026NatAs..10...92C}, while stellar magnetic activity traces the high-energy environment that may influence planetary atmospheres through atmospheric escape \citep{2012MNRAS.425.2931O,2021AJ....162..100C}. More recently, studies in a Galactic context have suggested that planet populations may also be linked to the chemical and dynamical evolution of the Milky Way \citep[e.g.,][]{2025ApJ...995...33C,2026ApJ..1001..176C}.

These correlations naturally lead to a deeper question: do planet-hosting stars themselves differ from otherwise similar stars without detected planets? Direct host–non-host comparisons offer a complementary way to probe the conditions associated with planet formation, beyond correlations between planet properties and host-star parameters alone. Previous studies have reported possible differences between host and non-host stars in chemical abundances, ages, and Galactic properties, but a consistent picture has not yet emerged \citep[e.g.,][]{2017A&A...599A..96S,2021AJ....162..100C,2022AJ....164..181U,2025ApJ...995...33C}. 

A key difficulty is that stellar properties are not independent. They are shaped jointly by stellar evolution and by the chemical evolution of the Milky Way. Thick-disk stars, for example, are generally older, more metal-poor, and more $\alpha$-enhanced than thin-disk stars, while recent studies have revealed a complex, non-monotonic age–metallicity relation even within the thin disk \citep{2013A&A...560A.109H,2018MNRAS.477.2326F,2019MNRAS.489.1742F,2022Natur.603..599X,2022MNRAS.512.2890L,2023MNRAS.523.1199S,2025NatCo..16.1581S}. Apparent differences in metallicity, elemental abundances, or activity may therefore reflect differences in age or Galactic population rather than signatures associated with planet formation itself. A meaningful comparison requires precise and homogeneous measurements of stellar ages, masses, and chemical abundances.

Such measurements have become increasingly available in recent years. Advances in astrometry, large-scale spectroscopy, and data-driven modeling have greatly improved the precision and homogeneity of stellar parameters. Gaia DR3 provides precise parallaxes and luminosities for vast stellar samples \citep{2023gaia}, while spectroscopic surveys such as LAMOST \citep{2012RAA....12.1197C}, GALAH \citep{2015MNRAS.449.2604D}, and APOGEE \citep{2017AJ....154...94M} provide homogeneous atmospheric parameters and detailed chemical abundances. In particular, the DD-Payne approach \citep{2019ApJS..245...34X} has enabled precise stellar parameters to be extracted from low-resolution LAMOST spectra, allowing accurate ages to be derived for large samples of subgiants and dwarfs when combined with Gaia constraints \citep{2022Natur.603..599X,2023ApJS..268...29S,2025ApJ...995...33C}. These developments have opened the door to a new generation of host--non-host studies based on well-characterized stellar samples.

In this work, we compare host stars with age–mass matched non-host stars after correcting for distribution-induced matching biases, and investigate differences in stellar chemistry, chromospheric activity, and Galactic birth environment. We further test whether these host–non-host differences depend on planet size by separately considering Earth-like planets, sub-Neptunes, and giant planets. The remainder of this paper is organized as follows. Section~\ref{sec:data} describes the sample selection, the determination of stellar parameters, and the matched-control methodology. Section~\ref{sec:results} presents the host–non-host comparisons in overall metallicity, elemental abundances, chromospheric activity, and stellar birth radius. Section~\ref{sec:summary} summarizes our main results and their implications for planet formation and evolution.

\section{Data and Methods} 
\label{sec:data}
\subsection{Sample Selection}

We adopt the quality-controlled Kepler--LAMOST--Gaia stellar sample compiled by \citet{2025ApJ...995...33C}, constructed from a cross-match of Kepler stars \citep{2020AJ....159..280B} with LAMOST DR9 DD-Payne \citep{2025ApJS..279....5Z} and Gaia DR3 \citep{2023gaia}, followed by spectroscopic, astrometric, abundance-quality, and photometric-completeness cuts. The resulting sample contains 44,148 stars with homogeneous stellar parameters and elemental abundances. Following \citet{2020AJ....159..279B}, we use the \texttt{evolstate} package to identify stellar evolutionary stages and exclude giant stars, leaving 34,488 dwarfs and subgiants for subsequent analysis. The construction of this parent sample is summarized in Table~\ref{tab:stellar_selection} (steps~1--5).

\begin{table*}
\centering
\scriptsize
\caption{Summary of the Sample Selection}
\label{tab:stellar_selection}
\begin{tabular}{clc}
\hline
Step & Selection Criteria & Number of Stars Remaining \\
\hline
1 & Kepler stellar sample \citep{2020AJ....159..280B} & 186,301 \\
2 & Cross-match with LAMOST DR9 DD-Payne ($<1.5^{\prime\prime}$) \citep{2025ApJS..279....5Z} & 76,058 \\
3 & Cross-match with Gaia DR3 \citep{2023gaia} & 74,581 \\
4 & Quality cuts (SNRG $\geq 20$, $\chi^2$ flags $\leq 3$, valid abundance flags, RUWE $\leq 1.2$, and available CDPP) & 44,148 \\
5 & Excluding giants & 34,488 \\
6 & Final stellar sample with ages ($\rm age < 15$~Gyr, relative error $<100\%$) & 28,383 \\
\hline
7 & \textit{From Step~6:} cross-match with Kepler DR25 planets with $P_{\rm orb} < 100$~days & 629 hosts (865 planets) \\
8 & \textit{From Step~6:} stars with derived $\log R^{+}_{\rm HK}$ & 17,020 \\
9 & \textit{From Step~6:} stars with derived $R_{birth}$ & 28,383 \\
\hline
\end{tabular}
\tablecomments{
Steps~1--4 follow the stellar sample selection of \citet{2025ApJ...995...33C}. 
Steps~7--9 define three \emph{parallel} subsamples drawn from the Step~6 sample (rather than sequential cuts) for analyses requiring stellar ages, $\log R^{+}_{\rm HK}$, and stellar birth radius, respectively. 
}
\end{table*}

\subsection{Stellar Fundamental Parameters}

Stellar fundamental parameters (luminosities, radii, masses, and ages) were derived following the methodology described in \citet{2026ApJ..1001..176C}. Briefly, stellar luminosities were computed using 2MASS $K$-band photometry, Gaia DR3 parallaxes corrected for the global zero-point offset \citep{2021A&A...649A...4L}, and extinction estimates from the high-resolution dust map of \citet{Wang_2025}. Stellar radii were then calculated from the derived luminosities and effective temperatures through the Stefan--Boltzmann relation.

\begin{figure*}[htbp]
  \centering
  \includegraphics[width=\textwidth]{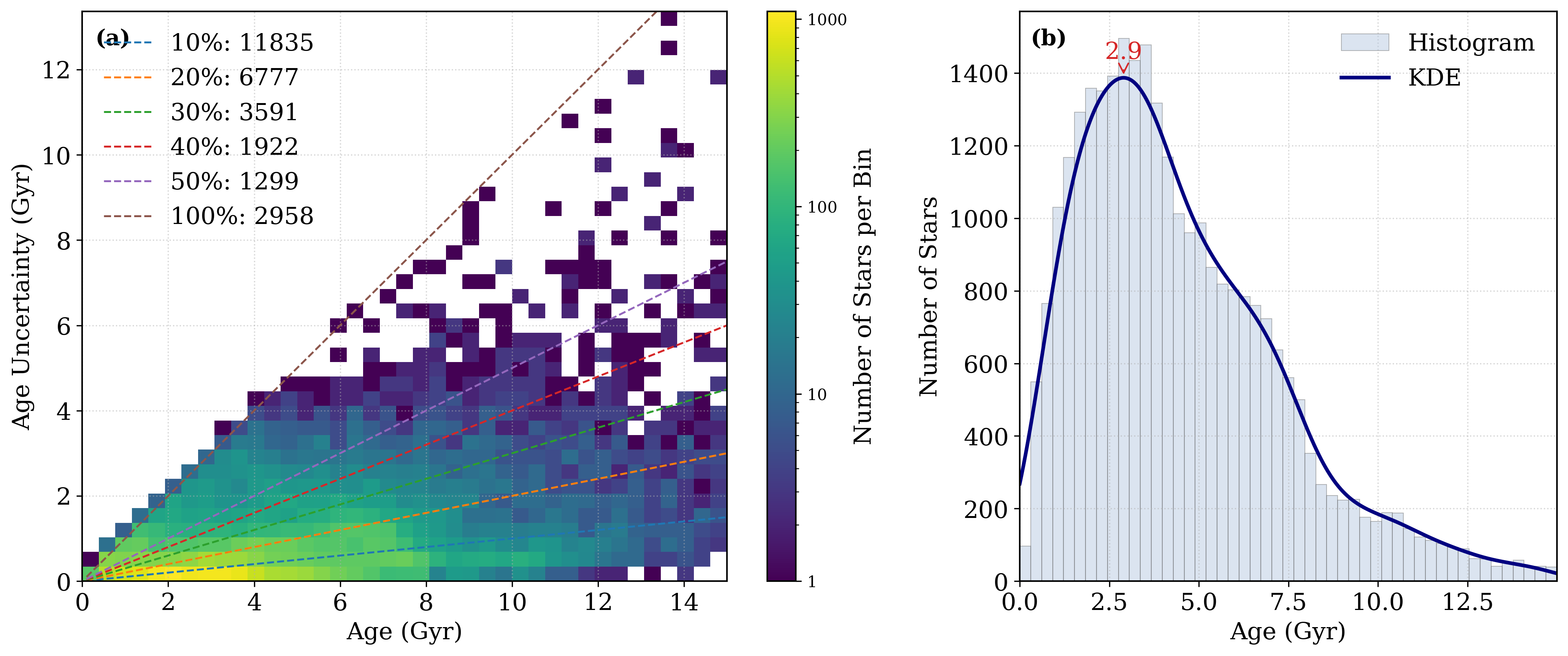}
  \caption{
  Age precision and distribution of 28,383 sample stars. 
  \textbf{(a)} Two-dimensional histogram of stellar age versus age uncertainty, with dashed lines indicating fractional age uncertainties of 10\%, 20\%, 30\%, 40\%, 50\%, and 100\%. 
  The colorbar indicates the number of stars per bin in logarithmic scale. 
  \textbf{(b)} Age histogram (gray bars) overlaid with a KDE-smoothed curve (black). Vertical dashed lines and arrows highlight the detected peaks in the age distribution.
  }
  \label{fig:age_distribution}
\end{figure*}

Stellar masses and ages were estimated using the Bayesian isochrone fitting framework, based on the formalism of \citet{2010A&A...522A...1K} and \citet{2010ApJ...710.1596B}. The fitting employed the observed stellar parameters $(T_{\rm eff}, L, \mathrm{[Fe/H]})$ and compared them with the Yonsei--Yale (YY) stellar isochrones \citep{2004ApJS..155..667D}. For each star, the model grid closest to its measured $[\alpha/{\rm Fe}]$ value was adopted, where $[\alpha/{\rm Fe}]$ was calculated as the error-weighted mean abundance of Mg, Si, Ca, and Ti from the LAMOST DR9 DD-Payne catalog. Stellar masses and ages were inferred from the posterior probability distributions. A Gaussian function was fitted to each posterior distribution, and the mean and standard deviation of the fitted Gaussian were adopted as the parameter estimate and its uncertainty, respectively. We retained stars with reliable age estimates satisfying ${\rm age}<15\,{\rm Gyr}$ and relative age uncertainties smaller than 100\%, resulting in a final sample of 28,383 stars, comprising 27,754 non-host stars and 629 host stars. Throughout this work, ``non-host'' refers to a star without a detected
Kepler planet and does not necessarily imply that the star is intrinsically
planet-free. Figure~\ref{fig:age_distribution} shows the precision and distribution of the derived stellar ages. The ages are generally well constrained, with a median uncertainty of 0.48~Gyr and a median fractional uncertainty of 13.1\%. The mean fractional uncertainty is 20.5\%, indicating that the age estimates are sufficiently precise for subsequent statistical analyses. Panel~(b) reveals a prominent age peak at $\sim2.9$~Gyr. The stellar masses are also well determined, with a median uncertainty of $0.015\,M_\odot$ (mean: $0.020\,M_\odot$).

\subsection{Stellar Activity}
Chromospheric activity was quantified using the $R^{+}_{\rm HK}$ index measured from the LAMOST DR9 spectra\footnote{https://www.lamost.org/dr9/v1.0/}. The Mount Wilson $S$-index was first derived from the Ca~II H\&K line cores and subsequently converted to $R^{+}_{\rm HK}$ using the calibration of \citet{2013A&A...549A.117M}. The $R^{+}_{\rm HK}$ index removes both the photospheric contribution and the basal chromospheric emission, providing a direct measure of magnetic activity-related chromospheric flux. Reliable $R^{+}_{\rm HK}$ values were obtained for 17,020 stars in the parent sample, including 16,617 non-host stars and 403 planet-host stars.

\subsection{Stellar Birth Radius}

The stellar birth radius, $R_{\rm birth}$, was estimated following the chemical--dynamical framework developed by \citet{2024MNRAS.535..392L}. Under the assumption of a radially linear interstellar medium metallicity profile, the birth radius can be expressed as

\begin{equation}
R_{\rm birth}=
\frac{{\rm [Fe/H]}-{\rm [Fe/H]}(R=0,\tau)}
{\nabla{\rm [Fe/H]}(\tau)},
\end{equation}

where $\tau$ is the stellar age, ${\rm [Fe/H]}(R=0,\tau)$ is the central ISM metallicity, and $\nabla{\rm [Fe/H]}(\tau)$ is the age-dependent radial metallicity gradient. Both quantities were obtained by interpolation of the calibration tables provided by \citet{2024MNRAS.535..392L}. The resulting $R_{\rm birth}$ values represent the inferred Galactocentric radii at which the stars were formed.

\subsection{Planet Radius}

\begin{figure*}[htbp]
  \centering
  \includegraphics[width=\textwidth]{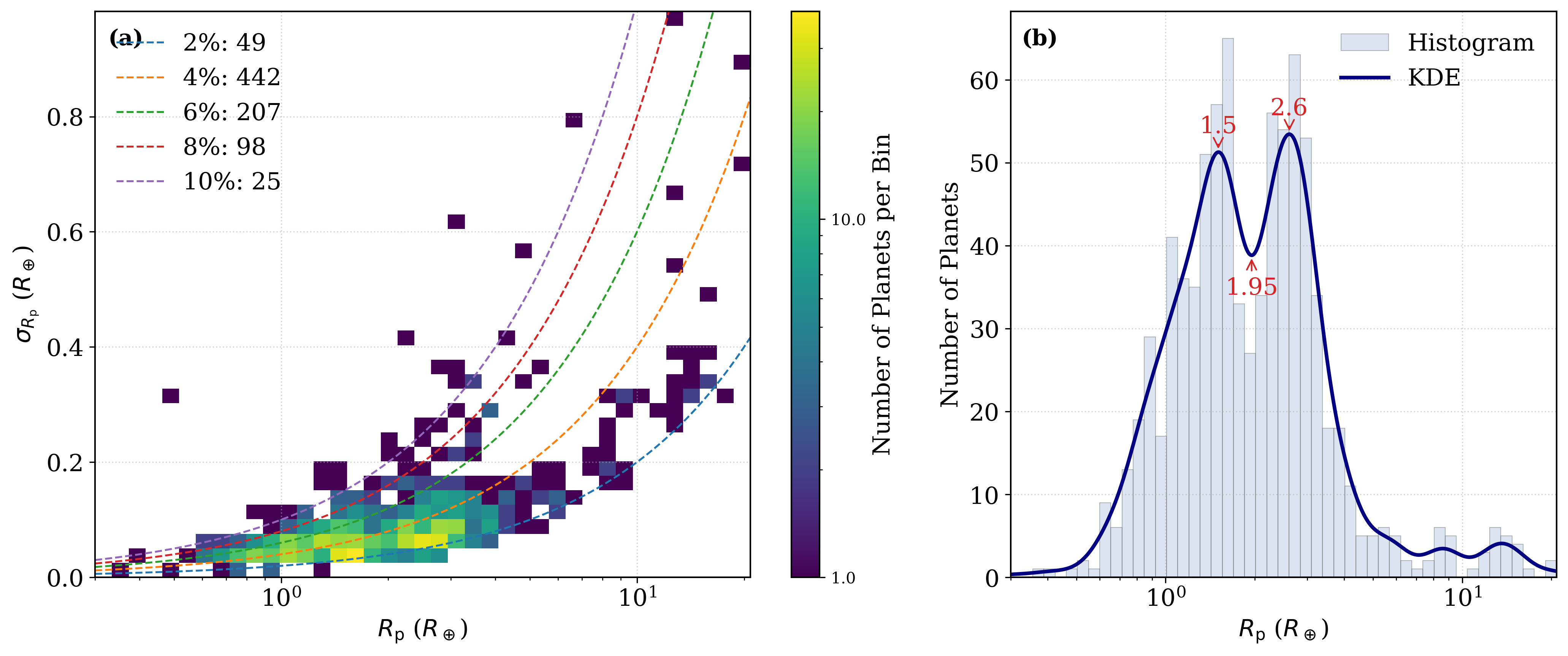}
  \caption{
  Radius precision and radius distribution for the 845 planets with measured radii in our sample.
  \textbf{(a)} Two-dimensional histogram of planetary radius versus radius uncertainty. Dashed lines indicate fractional radius uncertainties of 2\%, 4\%, 6\%, 8\%, and 10\%. The colorbar shows the number of planets per bin on a logarithmic scale.
  \textbf{(b)} Distribution of planetary radii shown as a histogram (light-blue bars) together with a kernel density estimate (KDE; navy curve). Red arrows mark the two dominant peaks at $R_{\rm p}\approx1.5,R_\oplus$ and $R_{\rm p}\approx2.6,R_\oplus$, as well as the intervening deficit near $R_{\rm p}\approx2,R_\oplus$, corresponding to the radius valley.
  }
  \label{fig:radius_distribution}
\end{figure*}

Planetary radii were recomputed using the updated stellar radii derived in this work together with the transit depths reported in the Kepler DR25 catalog \citep{Thompson_2018, k2pandc}. Among the 865 planets, 845 have available transit depths. Following the standard transit relation \citep[e.g.,][]{2025ApJ...993..233W}, the planetary radius is given by
\begin{equation}
R_{\rm p} = \sqrt{\Delta F}\, R_\star ,
\end{equation}
where $\Delta F$ is the transit depth and $R_\star$ is the stellar radius. The uncertainties in $R_{\rm p}$ were estimated by propagating the uncertainties in both the stellar radius and the transit depth. Figure~\ref{fig:radius_distribution} summarizes the precision and distribution of the derived planetary radii for the 845 planets. The radii are generally well constrained, with a median uncertainty of $0.069\,R_\oplus$ and a median fractional uncertainty of 3.6\% (mean: 4.3\%), as shown in panel~(a). Panel~(b) reveals a clear bimodal radius distribution, with peaks at $R_{\rm p}\approx1.5\,R_\oplus$ and $R_{\rm p}\approx2.6\,R_\oplus$, separated by a pronounced deficit near $R_{\rm p}\approx2\,R_\oplus$ corresponding to the well-known radius valley (e.g., \citet{2017AJ....154..109F, 2018AJ....155...89P}).

\subsection{Matched-control Analysis and Baseline Correction}

To investigate intrinsic differences between host stars and non-host stars, it is necessary to account for variations in their underlying stellar-parameter distributions. Stellar mass is closely related to the mass and evolution of the protoplanetary disk \citep{2013ApJ...771..129A,2016ApJ...831..125P}, while stellar age traces both the evolutionary stage of the system and the chemical enrichment history of the Galactic disk. Since many parameters analyzed in this work, including metallicity, elemental abundance ratios, chromospheric activity, and stellar birth radii depend on stellar age and mass, differences between host and non-host stars may arise from differences in their age--mass distributions rather than from planet hosting itself.

\begin{figure}[htbp]
\centering
\includegraphics[width=\linewidth]{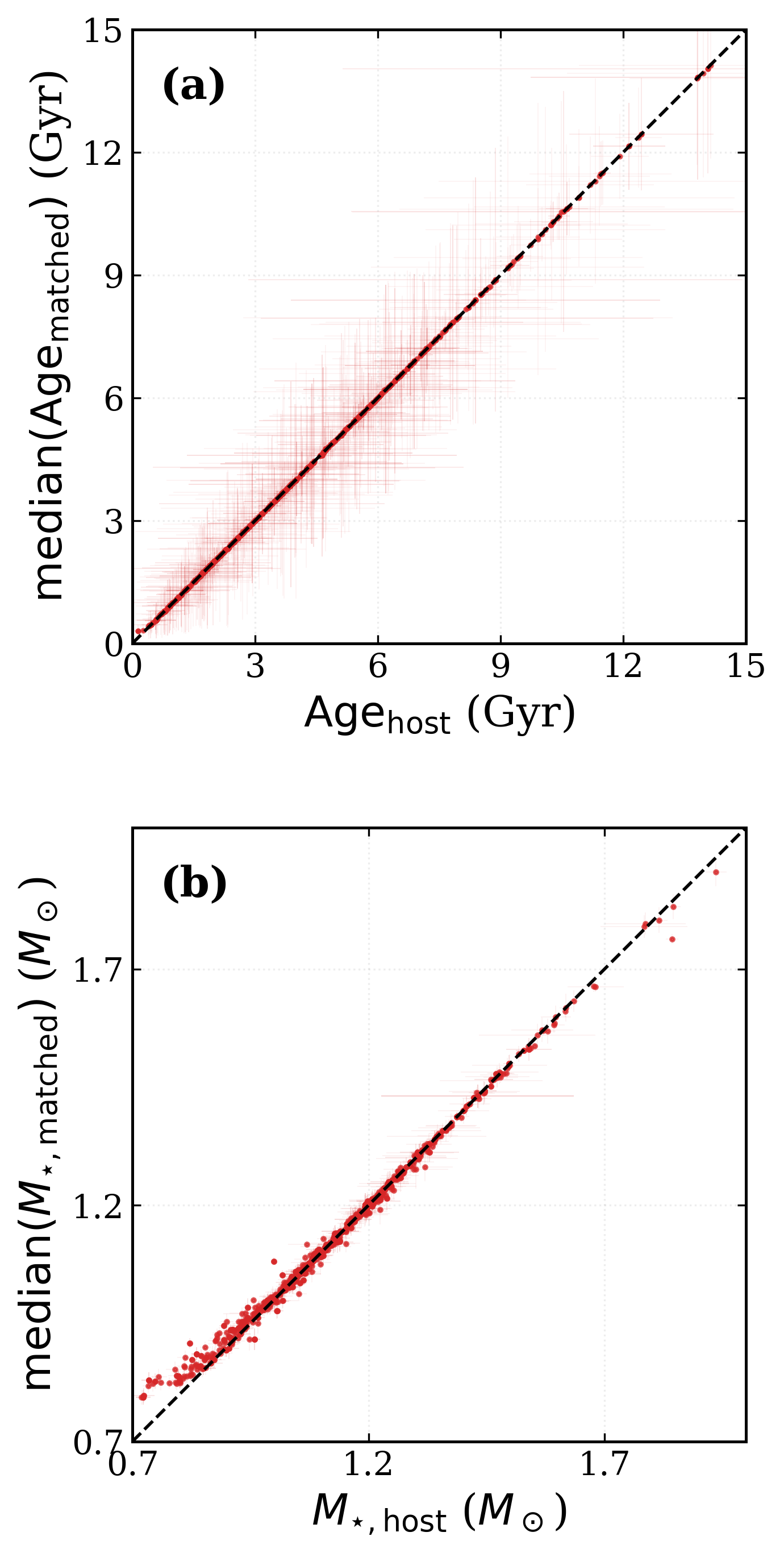}
\caption{
Quality of the age--mass matching procedure. 
Panel (a) compares the age of each planet-host star with the median age of its 50 matched non-host stars, while panel (b) shows the corresponding comparison for stellar mass. 
The dashed lines indicate the one-to-one relations. 
The error bars represent the reported uncertainties of the host stars and the median uncertainties of the corresponding matched non-host samples. 
}
\label{fig:matching_quality}
\end{figure}

To minimize these effects, we constructed an age--mass matched comparison sample. For each host star, we identified the 50 nearest non-host stars in the two-dimensional parameter space of stellar age and mass using the \texttt{NearestNeighbors} algorithm implemented in \texttt{scikit-learn}. Before matching, stellar age and mass were standardized by subtracting their sample means and dividing by their standard deviations. Nearest neighbors were then identified using the Euclidean distance in the normalized age--mass space. We adopted 50 matched non-host stars for each host star to provide a statistically robust control sample while preserving close agreement in age and mass. Figure~\ref{fig:matching_quality} illustrates the quality of the matching procedure by comparing the age and stellar mass of each planet-host star with the median values of its 50 matched non-host stars. The matched samples closely follow the one-to-one relations, demonstrating that the nearest-neighbor algorithm successfully identifies control samples with very similar age and mass distributions. However, as shown in Figure~\ref{fig:matching_quality}, the host--non-host differences are generally much smaller than the quoted uncertainties of the stellar parameters. Therefore, in the subsequent analysis, we evaluate host--non-host differences relative to the mean properties of the matched non-host stars rather than individual comparison stars. This approach reduces the impact of random scatter among individual control stars while providing a more stable estimate of the statistical differences between host stars and their matched control samples.

For a given parameter $X$, the raw difference between a host star and its matched non-host stars is defined as
\begin{equation}
\Delta X_{\rm raw}
=
X_{\rm host}
-
\langle X_{\rm match}\rangle ,
\end{equation}
where $\langle X_{\rm match}\rangle$ denotes the mean value of $X$ among the 50 matched non-host stars.

However, age--mass matching alone does not completely remove biases introduced by the intrinsic distribution of $X$. If the distribution of a parameter is non-uniform or bounded, stars near the low end of $X$ tend to be matched with stars of higher $X$, whereas stars near the high end of $X$ tend to be matched with stars of lower $X$. Consequently, a systematic trend in $\Delta X_{\rm raw}$ can arise even when the host and non-host samples are drawn from the same parent population.

To quantify this distribution-induced matching bias, we performed a Monte Carlo fake-host experiment using only the non-host sample. For each parameter $X$, non-host stars with valid measurements of age, mass, and $X$ were divided into eight bins spanning the central 5th--95th percentile range of the corresponding distribution. In each bin, we performed 1000 Monte Carlo realizations. In each realization, 100 non-host stars were randomly selected without replacement and treated as fake host stars. Each fake host star was matched to its 50 nearest non-host neighbors in the normalized age--mass space using the same procedure described above, excluding the fake host star itself from its matched sample. For each fake host star, we calculated
\begin{equation}
\Delta X_{\rm fake}
=
X_{\rm fake}
-
\langle X_{\rm match}\rangle .
\end{equation}

Repeating this procedure 1000 times yielded a distribution of mean offsets for each bin. The median of this distribution was adopted as the baseline matching bias, while the 16th–84th percentile range was used to characterize its uncertainty.

The final corrected difference is defined as
\begin{equation}
\Delta X_{\rm corr}
=
\Delta X_{\rm raw}
-
\Delta X_{\rm bias}(X),
\end{equation}
where $\Delta X_{\rm bias}(X)$ is the interpolated baseline matching bias evaluated at the host-star value of $X$. This correction accounts for systematic trends introduced by the matching procedure and by the intrinsic shape of the parameter distribution, thereby reducing distribution-induced biases and enabling a more reliable comparison between host stars and non-host stars.

\begin{figure*}[htbp]
    \centering
    \includegraphics[width=\textwidth]{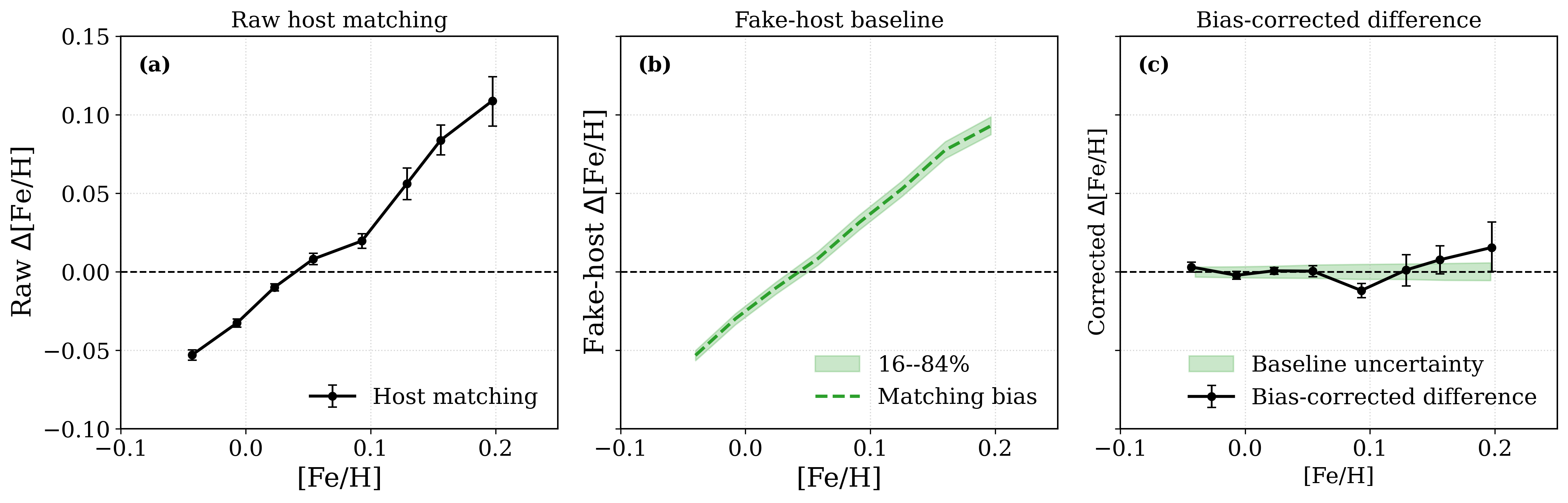}
    \caption{
    Illustration of the bias-correction procedure using [Fe/H] as an example.
    \textbf{(a)} Raw [Fe/H] difference between host stars and their age--mass matched non-host stars, defined as $\Delta$[Fe/H]${\rm raw}$. Points show the mean values in [Fe/H] bins, and error bars denote the 16th--84th percentile range estimated from 5000 bootstrap resamplings.
    \textbf{(b)} Baseline matching bias derived from the Monte Carlo fake-host experiment. The dashed curve shows the median bias, while the shaded region indicates the 16th--84th percentile range obtained from 1000 Monte Carlo realizations.
    \textbf{(c)} Bias-corrected [Fe/H] difference, $\Delta$[Fe/H]$_{\rm corr}$, obtained after subtracting the interpolated baseline matching bias from the raw host--non-host difference. Black points and error bars represent the mean corrected offsets and their 16th--84th percentile uncertainties estimated from 5000 bootstrap resamplings. The green shaded region is the same baseline uncertainty range shown in panel~(b).}
    \label{fig:method_bias_correction}
\end{figure*}

Figure~\ref{fig:method_bias_correction} illustrates the bias-correction procedure using [Fe/H] as an example. Panel~(a) shows the raw host–non-host [Fe/H] difference as a function of host-star [Fe/H] after age–mass matching. A clear systematic trend is present, with metal-poor host stars tending to exhibit negative offsets and metal-rich host stars tending to exhibit positive offsets. However, panel~(b) demonstrates that a very similar trend is recovered from the Monte Carlo fake-host experiment performed using only non-host stars, indicating that much of the apparent [Fe/H] dependence originates from the matching procedure and from the intrinsic shape of the [Fe/H] distribution itself. After subtracting the interpolated baseline matching bias, the corrected differences shown in panel~(c) are substantially reduced and become more consistent with zero across the [Fe/H] range. This example demonstrates that a substantial fraction of the raw host--non-host differences can arise from distribution-induced matching biases. In particular, the [Fe/H] trend reported in \citet{2025ApJ...995...33C}, in which host stars appeared more metal-poor than non-host stars at ${\rm [Fe/H]}<-0.2$, is significantly reduced after accounting for the intrinsic shape of the [Fe/H] distribution. Correcting for these biases is therefore essential to obtain a more reliable comparison between host and non-host stars.

An additional limitation of the present analysis is that stars classified as non-hosts are more precisely stars without detected Kepler planets. Because of geometric transit probability and observational incompleteness, some of these stars may host planets that remain undetected. Their inclusion in the comparison sample would tend to dilute intrinsic host--non-host differences and may partly account for the modest amplitudes and statistical significance of the corrected offsets reported below. We therefore interpret the present results as differences between stars with detected planets and stars without detected planets.

\section{Results}
\label{sec:results}
In this section, we investigate whether host stars differ systematically from matched non-host stars in several stellar parameters that are expected to influence planet formation and evolution, including stellar overall metallicity, elemental abundance ratios, chromospheric magnetic activity ($\log R^+_{\rm HK}$), and stellar birth radius ($R_{\rm birth}$). To explore possible dependencies on planet size, we divide the planet sample into three radius regimes: 443 Earth-like planets ($R_{\rm p}<2,R_\oplus$), 331 sub-Neptunes ($2\leq R_{\rm p}<4,R_\oplus$), and 71 giant planets ($R_{\rm p}\geq4,R_\oplus$). This classification is motivated by the presence of the radius valley near $R_{\rm p}\approx2,R_\oplus$, which separates two dominant populations of close-in planets and is thought to reflect distinct evolutionary pathways \citep{2013ApJ...776....2L,2013ApJ...775..105O,2014ApJ...795...65J,2016ApJ...831..180C,2018MNRAS.476..759G,2019MNRAS.487...24G,2020MNRAS.493..792G}.

\subsection{Host--Non-host Differences in Overall Metallicity}
Stellar metallicity is a fundamental parameter for planet formation because it regulates the amount of solid material available in protoplanetary disks. Numerous studies have shown that giant-planet occurrence increases strongly with host-star metallicity \citep[e.g.,][]{2004A&A...415.1153S,2005ApJ...622.1102F,2010PASP..122..905J}, while recent work has suggested a metallicity threshold for the formation of short-period super-Earths \citep{2024AJ....168..128B}. Moreover, $\alpha$-enrichment may facilitate the formation of large planets even in metal-poor environments \citep{2026ApJ...998..301G}. These results suggest that planet formation depends on both iron abundance and $\alpha$-element enrichment. We therefore investigate the differences in the overall metallicity $Z$ between host and non-host stars, since $Z$ incorporates the contributions of both [Fe/H] and [$\alpha$/Fe].

\begin{figure*}[htbp]
    \centering
    \includegraphics[width=\textwidth]{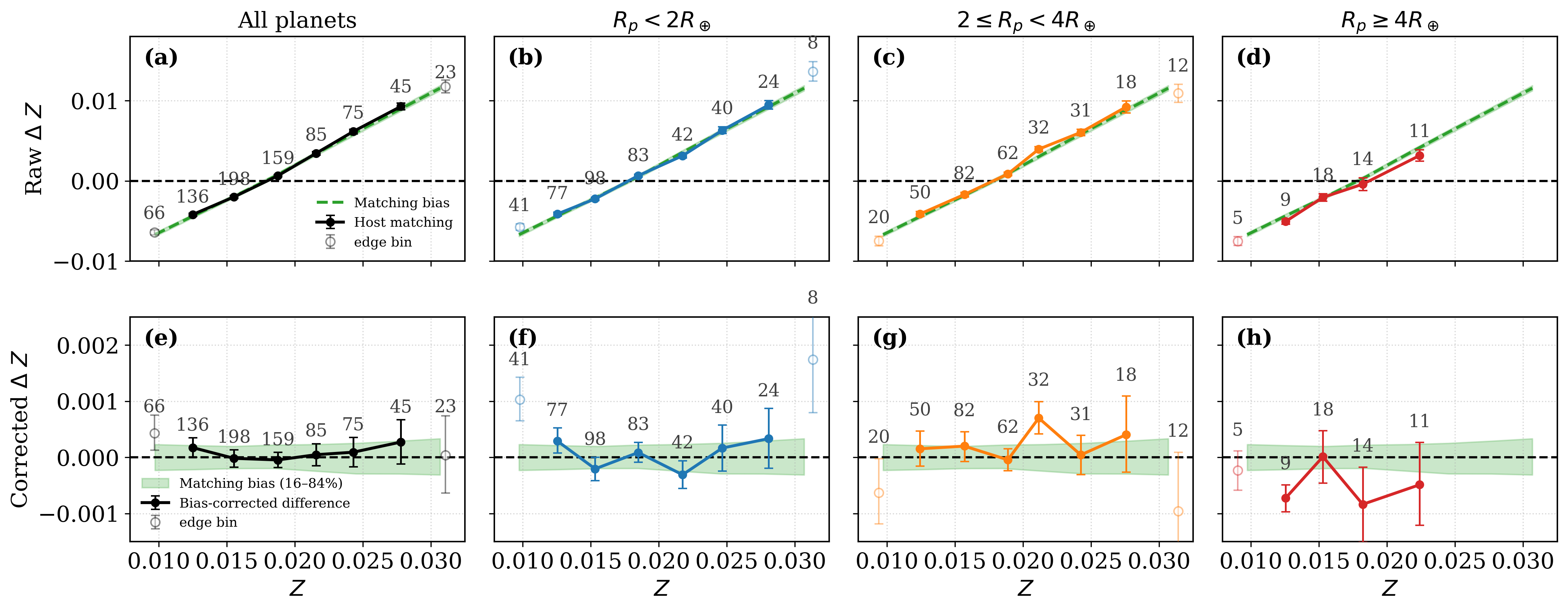}
    \caption{
    Host--non-host comparison in overall metallicity $Z$.
    Top row: raw metallicity difference, $\Delta Z_{\rm raw}$, between host stars and their age--mass matched non-host stars. 
    Bottom row: corrected metallicity difference, $\Delta Z_{\rm corr}$, after subtracting the fake-host baseline matching bias. 
    Columns show results for all planets, Earth-like planets ($R_{\rm p}<2\,R_\oplus$), sub-Neptunes ($2\leq R_{\rm p}<4\,R_\oplus$), and giant planets ($R_{\rm p}\geq4\,R_\oplus$), respectively. 
    Green dashed curves and shaded regions indicate the median fake-host matching bias and its 16th--84th percentile range. 
    Points and error bars show the binned host-star offsets and their 16th--84th percentile bootstrap uncertainties. 
    Open symbols mark edge bins, and numbers indicate the number of planets in each bin. Because the edge bins are affected by both small-number statistics and boundary effects arising from the finite parameter coverage and matching-bias correction, they are displayed for completeness but are not used in the scientific interpretation of the observed trends.
    }
    \label{fig:z_matching}
\end{figure*}

Figure~\ref{fig:z_matching} presents the host--non-host comparison in the overall metallicity $Z$. For the full planet sample, the corrected $Z$ offsets remain close to zero over the entire $Z$ range, indicating no significant difference in $Z$ between host stars and their age--mass matched non-host stars. A similar behavior is observed for both the Earth-like and sub-Neptune subsamples, whose corrected offsets are generally consistent with zero within the uncertainties. The giant-planet subsample exhibits systematically negative offsets across most $Z$ bins. To quantify the amplitude of this effect, we converted the measured $\Delta Z$ values into equivalent $\Delta{\rm [Fe/H]}$ offsets assuming solar abundance ratios. For the four giant-planet bins, the corrected $Z$ offsets correspond to $\Delta{\rm [Fe/H]}=-0.027$, $+0.000$, $-0.022$, and $-0.010$ dex at ${\rm [Fe/H]}=-0.127$, $-0.037$, $+0.043$, and $+0.139$, respectively. Thus, except for one bin that is consistent with zero, giant-planet hosts are systematically more metal-poor than their age–mass matched non-host stars. The effect is strongest at the metal-poor end, reaching $\sim0.03$ dex at ${\rm [Fe/H]}\approx-0.13$, and decreases to only $\sim0.01$ dex at ${\rm [Fe/H]}\approx+0.14$. Although the statistical significance is limited by the relatively small number of giant planets in our sample, this result is noteworthy because giant planets are generally thought to preferentially form around metal-rich stars \citep[e.g.,][]{2014Natur.509..593B,2014ApJ...789L...3D,2015AJ....149..143F,2018AJ....155...89P,2021AJ....161..114S,2022AJ....164...60S}. Our results indicate that giant-planet hosts tend to have lower overall metallicities than age--mass matched non-host stars, with the largest deficit occurring in the metal-poor regime.

Overall, we do not find compelling evidence that planet-host stars possess systematically different overall metallicities from non-host stars once differences in stellar age and mass are taken into account. This suggests that the strong dependence of planet occurrence on stellar metallicity does not necessarily translate into a corresponding metallicity difference between host and non-host stars. The only notable deviation is observed for giant-planet hosts, which tend to exhibit lower overall metallicities than their age--mass matched non-host stars, although this result remains tentative because of the limited size of the giant-planet sample.

\subsection{Host--Non-host Differences in Elemental Abundance Ratios}
Stellar elemental abundances provide important constraints on planet formation because they regulate both the solid content of protoplanetary disks and the chemical composition of planetary building blocks. Here we focus on four key elements, C, O, Mg, and Si, which play important roles in the formation and evolution of planetary systems. These elements regulate the solid composition of protoplanetary disks and are closely linked to the composition and internal structure of planets \citep{2016ApJ...831...20B,2023MNRAS.524.6295P,2026ApJ...998..301G,2026ApJ..1003L..16Y}.

\begin{figure*}[htbp]
    \centering
    \includegraphics[width=\textwidth]{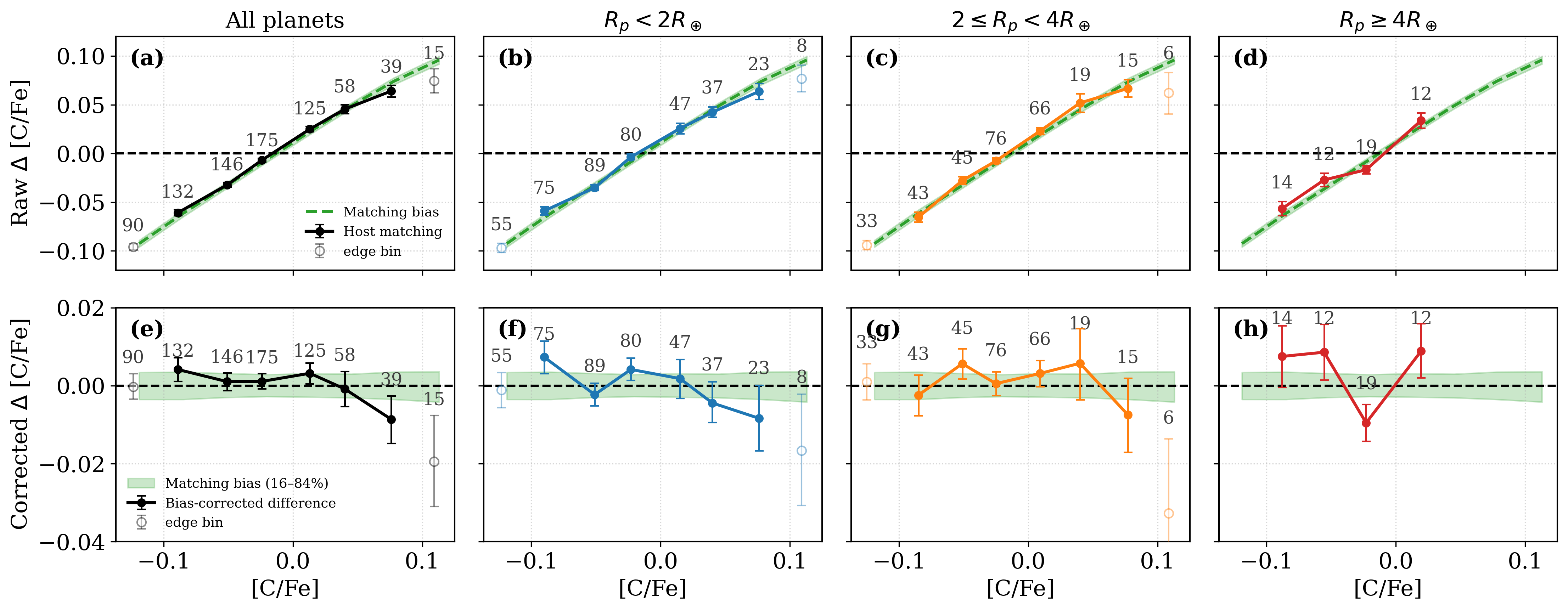}
    \caption{
    Same as Figure~\ref{fig:z_matching}, but for [C/Fe].
    }
    \label{fig:c_matching}
\end{figure*}

\begin{figure*}[htbp]
    \centering
    \includegraphics[width=\textwidth]{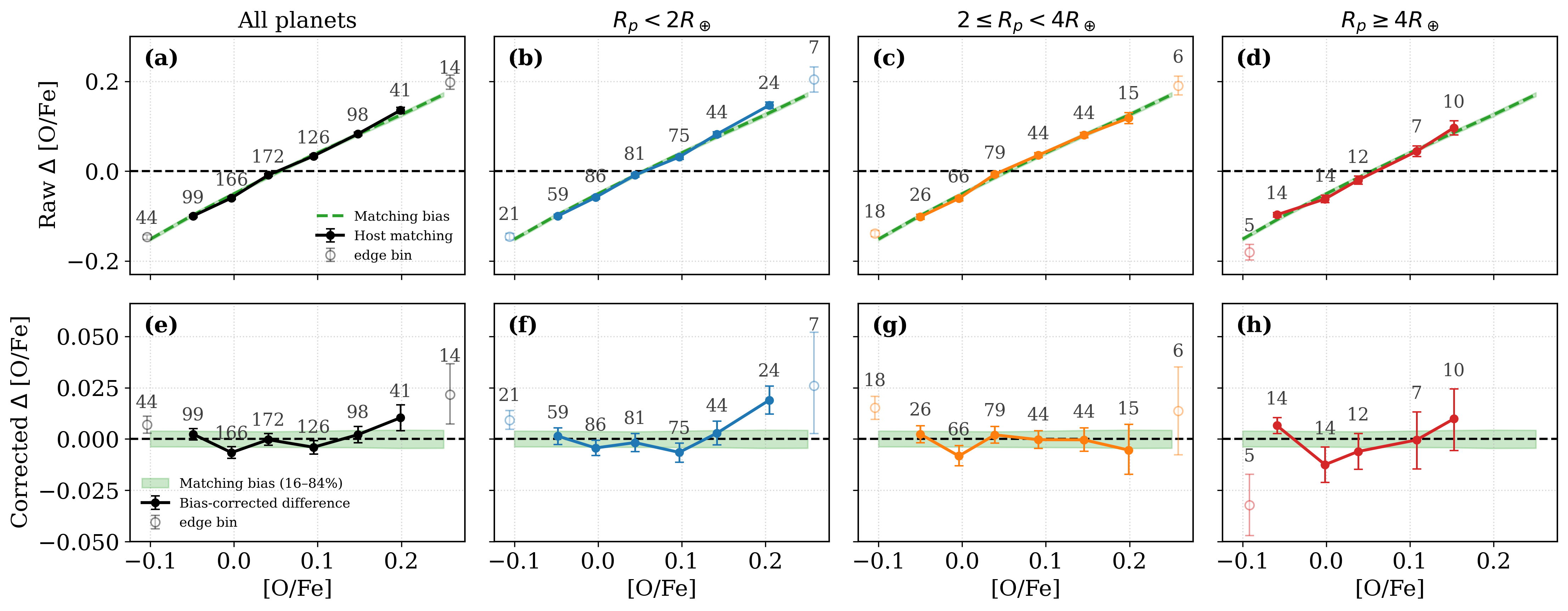}
    \caption{
    Same as Figure~\ref{fig:z_matching}, but for [O/Fe].
    }
    \label{fig:o_matching}
\end{figure*}

\begin{figure*}[htbp]
    \centering
    \includegraphics[width=\textwidth]{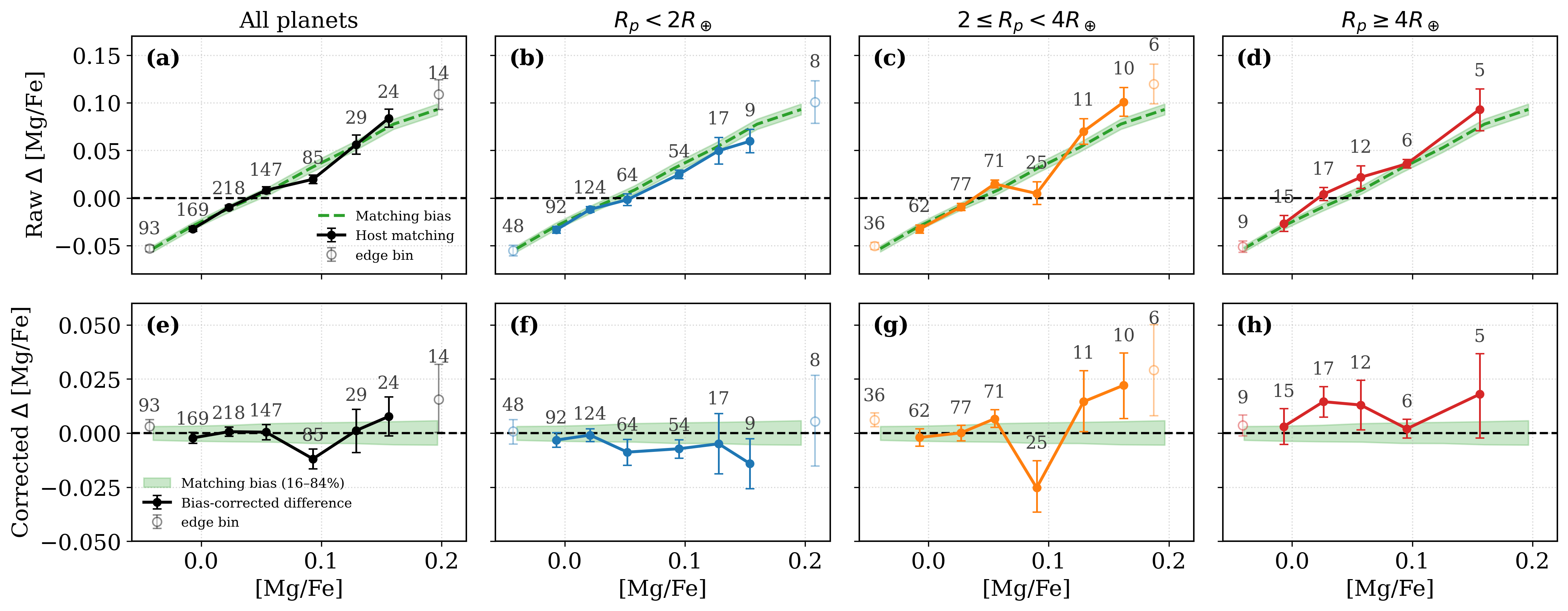}
    \caption{
    Same as Figure~\ref{fig:z_matching}, but for [Mg/Fe].
    }
    \label{fig:mg_matching}
\end{figure*}

\begin{figure*}[htbp]
    \centering
    \includegraphics[width=\textwidth]{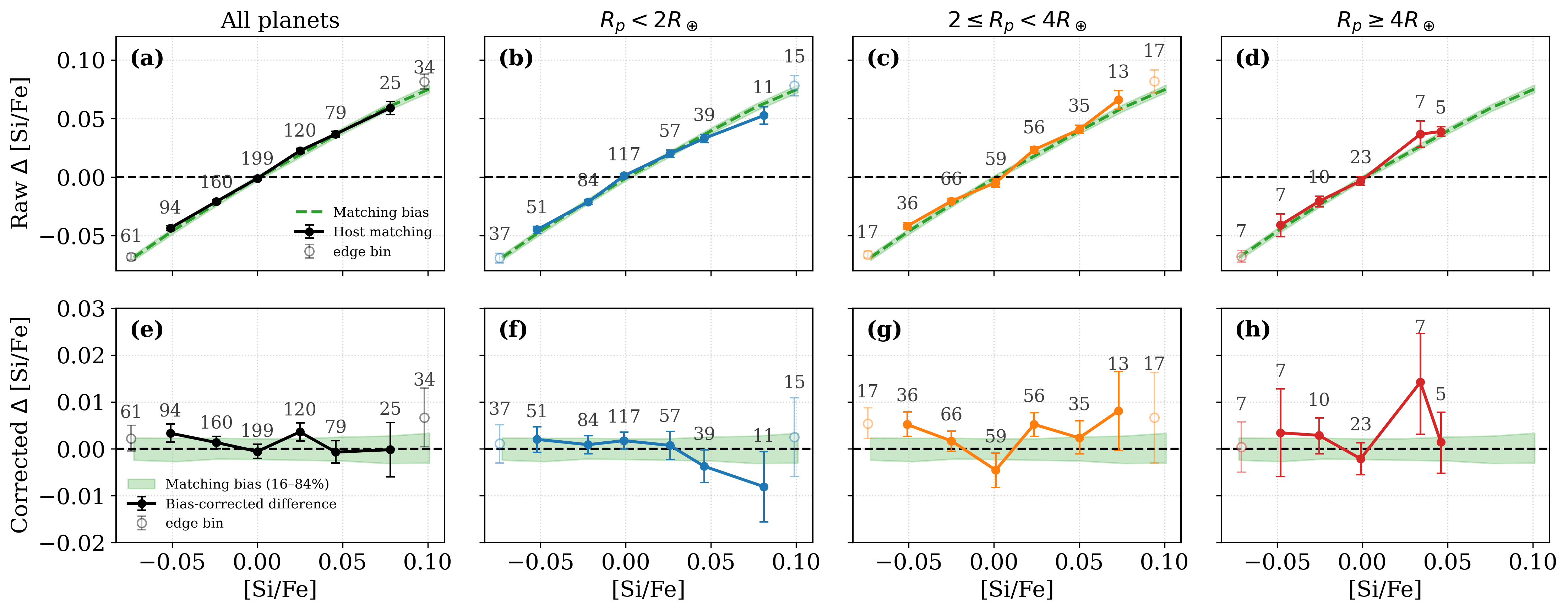}
    \caption{
    Same as Figure~\ref{fig:z_matching}, but for [Si/Fe].
    }
    \label{fig:si_matching}
\end{figure*}

Figures~\ref{fig:c_matching}--\ref{fig:si_matching} show host--non-host abundance differences for C, O, Mg, and Si. For the full planet sample, all four elements exhibit relatively small corrected offsets that are generally consistent with zero, indicating no strong abundance differences between host stars and age--mass matched non-host stars. This result differs from several previous studies that reported volatile-element depletion or refractory-element enhancement in planet-hosting stars \citep[e.g.,][]{2009ApJ...704L..66M,2009A&A...508L..17R,2017A&A...599A..96S}. The absence of abundance differences after controlling for stellar age and mass suggests that some previously reported host--non-host abundance signatures may be influenced by differences in stellar populations or underlying parameter distributions.

Although the full planet sample shows no significant abundance differences relative to age–mass matched non-host stars, distinct trends emerge when the sample is divided by planet radius. Carbon shows little difference between Earth-like and sub-Neptune hosts, indicating that small-planet hosts do not exhibit systematic carbon abundance variations relative to age–mass matched non-host stars. In contrast, giant-planet hosts show a slight overall tendency toward higher [C/Fe]. While the abundance differences are modest, three of the four [C/Fe] bins display positive host–non-host offsets, including both bins with [C/Fe] $\leq -0.05$, which consistently exhibit enhanced [C/Fe] relative to matched non-host stars. This behavior is qualitatively consistent with the results of \citet{2017A&A...599A..96S}, who reported that planet-host stars are relatively carbon-rich. Oxygen exhibits a different pattern. At [O/Fe] $< 0.1$, both Earth-like and sub-Neptune hosts are consistent with their matched non-host stars. Above [O/Fe] $\sim0.1$, however, their behaviors diverge. Earth-like hosts display increasingly positive abundance offsets with increasing [O/Fe], reaching a maximum of about 0.02 dex, whereas sub-Neptune hosts show a modest decline toward slightly negative offsets. Since C and O dominate the volatile inventory of protoplanetary disks and jointly regulate the C/O ratio \citep{2014prpl.conf..363P,2013pccd.book.....G}, these results suggest that volatile-element abundances may influence the formation pathways of different planet populations. In particular, enhanced oxygen abundances favor silicate-rich planetesimal formation and may therefore promote the production of rocky planets \citep{2005astro.ph..4214K,2014ApJ...787...81M}.

The clearest radius-dependent differences are observed for the refractory elements Mg and Si. For [Mg/Fe] $< 0.1$ and for [Si/Fe] $< 0$, both Earth-like and sub-Neptune hosts are generally indistinguishable from matched non-host stars. At higher abundances ([Mg/Fe] $\gtrsim 0.1$ and [Si/Fe] $\gtrsim 0.02$), however, the two populations begin to diverge. Earth-like hosts tend to show increasingly negative host–non-host offsets toward higher Mg and Si abundances, whereas sub-Neptune hosts appear to exhibit the opposite tendency, with relatively higher corrected offsets at the high-abundance end. For Mg, the sub-Neptune sequence includes a pronounced negative offset near [Mg/Fe]$\sim0.1$, although the offsets become more positive toward the most Mg-rich bins. Because the highest-abundance bins contain relatively few systems, these abundance-dependent trends should be interpreted with some caution. Nevertheless, similar opposite tendencies are seen for both Mg and Si, lending qualitative support to the possibility that Earth-like and sub-Neptune hosts respond differently at the high-abundance end. Giant-planet hosts display a mild tendency toward Mg enhancement, with positive offsets in most abundance bins, although the corresponding Si trend is less evident. Because Mg and Si are the principal rock-forming elements that regulate silicate mineralogy and planetary interior composition through the Mg/Si ratio \citep{2012ApJ...747L...2C,2016ApJ...831...20B}, these contrasting tendencies may suggest that Earth-like and sub-Neptune planets form in somewhat different chemical environments or inherit systematically different rocky building blocks. This interpretation is consistent with recent evidence that Earth-like planets and sub-Neptunes represent dynamically distinct populations with different evolutionary histories \citep{2026ApJ..1003L..16Y}.

In summary, host and non-host stars show remarkably similar abundance distributions overall, with typical host–non-host abundance offsets of less than $\sim0.05$ dex. This suggests that any host–non-host differences in stellar chemistry are generally subtle once stellar age and mass are taken into account. Nevertheless, separating planets by radius reveals patterns that are not apparent in the full sample. The most notable examples are found for O, Mg, and Si, where Earth-like and sub-Neptune hosts tend to exhibit different abundance-dependent trends relative to their matched non-host stars. Although these trends are modest and the highest-abundance bins contain relatively few systems, the contrasting abundance-dependent trends observed at the high-abundance end for Earth-like and sub-Neptune hosts in O, Mg, and Si suggests that subtle differences in stellar composition may still play a role in shaping the formation and subsequent evolution of different planetary populations.

\subsection{Host--Non-host Differences in Stellar Magnetic Activity}

\begin{figure*}[htbp]
    \centering
    \includegraphics[width=\textwidth]{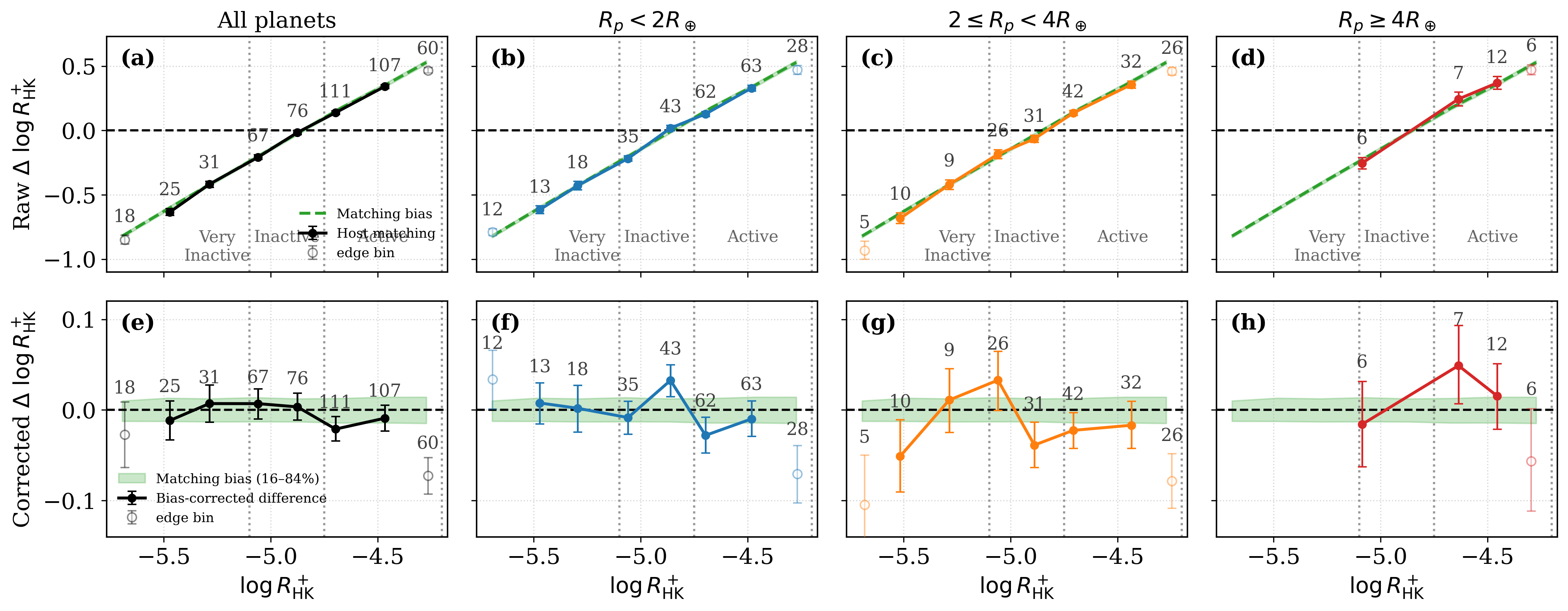}
    \caption{Same as Figure~\ref{fig:z_matching}, but for the chromospheric activity indicator $\log R^+_{\rm HK}$. Vertical dotted lines mark the approximate boundaries between the very inactive, inactive, and active activity regimes.}
    \label{fig:rhk_matching}
\end{figure*}

Stellar magnetic activity traces the high-energy radiative and magnetically driven environment experienced by planetary atmospheres, and is therefore expected to influence the evolution of close-in planets through atmospheric escape and star--planet interaction processes \citep{2012MNRAS.425.2931O,2021AJ....162..100C}. Figure~\ref{fig:rhk_matching} compares the chromospheric activity levels of host and non-host stars using $\log R^+_{\rm HK}$. Earth-like planet hosts show corrected offsets that are generally consistent with zero across most activity regimes, indicating no significant activity difference relative to age--mass matched non-host stars. In contrast, sub-Neptune hosts exhibit a weak tendency toward negative corrected offsets in several bins. This behavior is not present over the full activity range, but becomes apparent at $\log R^+_{\rm HK}\gtrsim-5$, where the offsets reach approximately $-0.04$ dex. These results suggest that relatively active sub-Neptune hosts may be slightly less magnetically active than comparable non-host stars. The giant-planet subsample does not exhibit a robust global trend because of the limited number of systems.

The modest tendency toward lower activity levels among relatively active sub-Neptune hosts may provide an additional clue to their evolutionary history. Since sub-Neptunes are expected to retain substantial volatile envelopes, a less intense high-energy irradiation environment could reduce atmospheric escape and thereby increase the likelihood of preserving primordial gaseous envelopes. This interpretation is broadly consistent with photoevaporation models, in which X-ray and extreme-ultraviolet irradiation associated with stellar activity drives atmospheric mass loss \citep{2013ApJ...775..105O,2021MNRAS.508.5886R,https://doi.org/10.48550/arxiv.2606.27107}. It is also qualitatively consistent with the scenario proposed by \citet{doi:10.1126/science.adu3916}, in which mini-Neptunes predominantly undergo relatively quiescent secular evolution and are therefore more likely to retain their primordial gaseous envelopes. However, given the small amplitude of the observed activity offsets, this interpretation should be regarded as suggestive rather than conclusive.

\subsection{Host--Non-host Differences in Stellar Birth Radius}

\begin{figure*}[htbp]
    \centering
    \includegraphics[width=\textwidth]{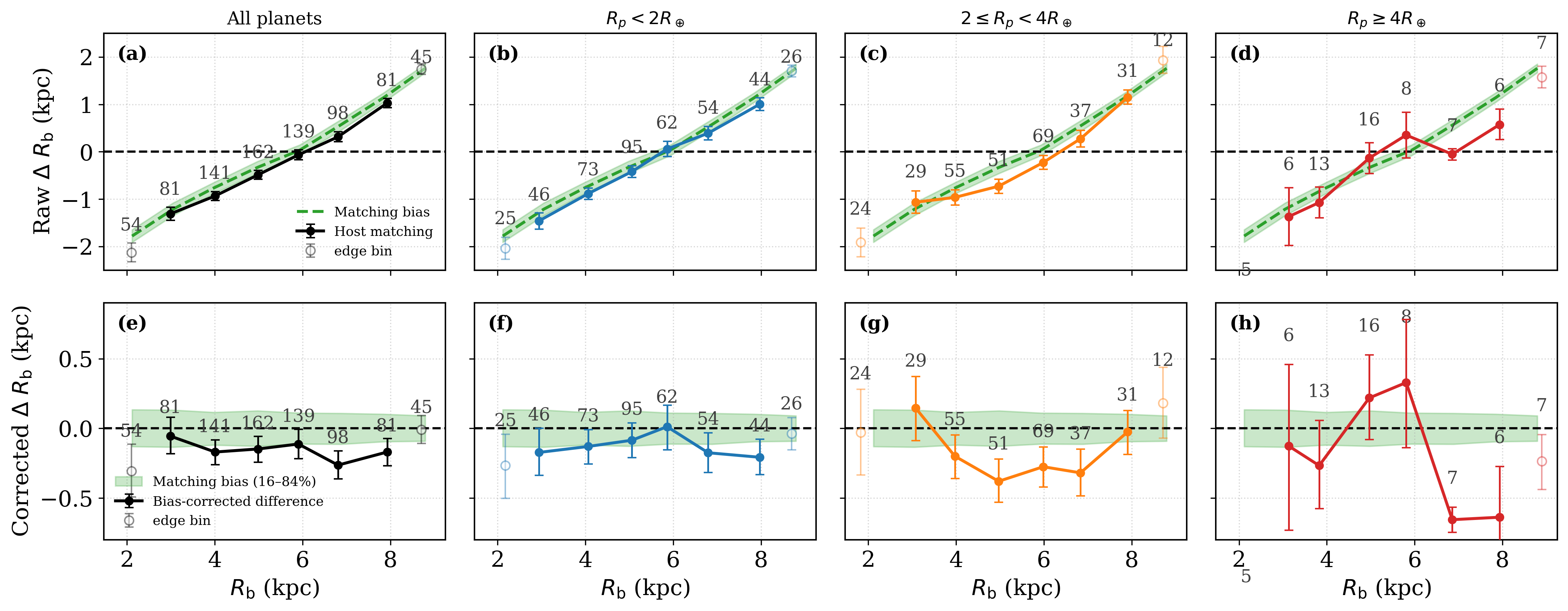}
    \caption{Same as Figure~\ref{fig:z_matching}, but for the stellar birth radius $R_{\rm b}$.}
    \label{fig:Rb_matching}
\end{figure*}

The stellar birth radius, $R_{\rm b}$, provides a useful tracer of the Galactic environment in which planetary systems formed. Because the chemical enrichment history, gas surface density, and star-formation conditions vary systematically across the Galactic disk, stars born at different radii may have experienced distinct protoplanetary disk environments. 

Figure~\ref{fig:Rb_matching} compares the birth radii of planet-hosting stars and age–mass matched non-host stars. For the full planet sample, the corrected host–non-host differences are generally small, with most bins remaining within $\sim0.2$–$0.3$ kpc of zero. Earth-like planet hosts show a similar behavior, with corrected offsets that are consistent with zero across nearly the entire birth-radius range. This suggests that Earth-like planet hosts do not preferentially originate from different Galactic environments than comparable non-host stars once age and stellar mass are controlled. In contrast, the sub-Neptune population exhibits a more coherent trend. At small birth radii ($R_{\rm b}\lesssim3$ kpc), the corrected offsets are consistent with zero. However, over the intermediate range $R_{\rm b}\sim4$–$7$ kpc, sub-Neptune hosts show systematically negative offsets of approximately $-0.2$ to $-0.4$ kpc relative to matched non-host stars. This indicates that stars hosting sub-Neptunes were, on average, born closer to the Galactic center than otherwise similar stars without detected planets. At larger birth radii, the trend weakens and becomes consistent with zero within the uncertainties. The giant-planet subsample shows larger fluctuations, with both positive and negative deviations appearing across different bins. However, the number of giant-planet hosts is small ($N\lesssim16$ per bin), resulting in substantially larger uncertainties. No robust systematic trend can therefore be established for the giant-planet population.

The tendency for sub-Neptune hosts to originate from smaller birth radii may provide an additional clue to their formation and evolutionary history. Stars born in the inner Galactic disk generally formed, on average, in environments characterized by higher gas densities and more rapid chemical enrichment, conditions that may favor the formation of massive rocky cores and the subsequent accretion of gaseous envelopes. Combined with the abundance results presented above, which show enhanced Mg and Si abundances among sub-Neptune hosts at the high-abundance end, and the lower chromospheric activity levels of sub-Neptune hosts, the birth-radius trend suggests that sub-Neptunes preferentially formed in environments that were both chemically enriched and conducive to retaining primordial volatile envelopes. In contrast, the absence of a corresponding birth-radius signal among Earth-like hosts indicates that the formation and evolution of rocky planets are likely less sensitive to Galactic birth environment than those of sub-Neptunes.

\subsection{Host--Non-host Comparisons for Hot and Longer-period Jupiter Systems}

To examine whether close-in and longer-period Jupiter-sized planets are associated with different stellar properties, we compare their host--non-host differences separately. Following the giant-planet radius interval adopted by \citet{Zhu2021}, we select planets with $8R_\oplus \leq R_{\rm p}<20R_\oplus$. We identify 17 planets with $P<10$ days as hot Jupiters and the remaining 15 planets with $P\geq10$ days as longer-period Jupiters. For each host star, we apply the same age--mass matching and fake-host baseline correction used in the main analysis.

\begin{figure*}[htbp]
\centering
\includegraphics[width=\textwidth]
{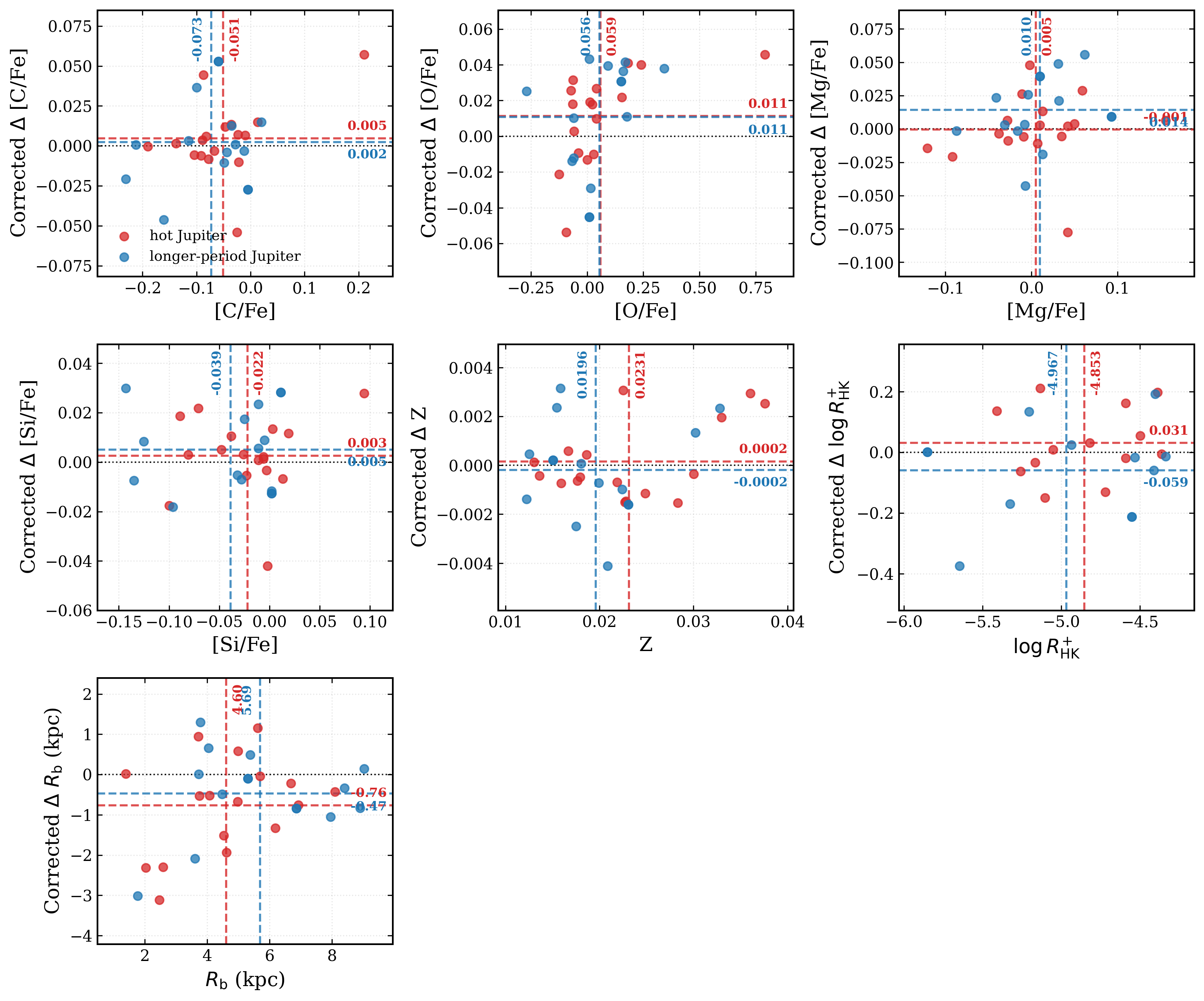}
\caption{
Corrected host--non-host differences in stellar properties for hot-Jupiter and longer-period Jupiter systems. We select Jupiter-sized planets with $8R_\oplus \leq R_{\rm p}<20R_\oplus$ and divide them into hot Jupiters with $P<10$ days and longer-period Jupiters with $P\geq10$ days. Red and blue points represent the hot-Jupiter and longer-period Jupiters samples, respectively. For each stellar parameter $X$, the horizontal axis shows the host-star value, while the vertical axis shows the bias-corrected difference,
$\Delta X_{\rm corrected}$, between the host star and the mean value of its 50 age--mass-matched non-host stars. The red and blue vertical dashed lines mark the mean host-star parameter values of the two samples, whereas the corresponding horizontal dashed lines indicate their mean corrected host--non-host differences. The numerical values adjacent to the dashed lines give the corresponding sample means. The black dotted line denotes $\Delta X_{\rm corrected}=0$.
}
\label{fig:hot_long_jupiter}
\end{figure*}

Figure~\ref{fig:hot_long_jupiter} shows that hot-Jupiter hosts and longer-period Jupiter hosts, exhibit broadly similar corrected host--non-host differences in [C/Fe], [O/Fe], [Mg/Fe], and [Si/Fe]. The two populations overlap substantially, and their mean offsets are generally small compared with the star-to-star scatter. The most suggestive, although still tentative, contrasts appear in the overall metallicity, chromospheric activity, and stellar birth radius. Hot-Jupiter hosts have a higher mean overall metallicity than longer-period Jupiter hosts ($Z=0.0231$ versus $0.0196$), with mean corrected offsets of $+0.0002$ and $-0.0002$, respectively. This trend is qualitatively consistent with previous evidence that eccentric hot Jupiters preferentially orbit metal-rich stars \citep{2013ApJ...767L..24D}. Hot-Jupiter hosts also show a higher mean activity level, with $\langle\log R^{+}_{\rm HK}\rangle=-4.853$ and a corrected mean offset of $+0.031$, compared with $-4.967$ and $-0.059$ for longer-period Jupiter hosts. This may suggest that chromospheric activity is also related to the presence or orbital configuration of close-in giant
planets, potentially through their influence on stellar rotation \citep{2007MNRAS.374L..42C,2014MNRAS.442.2081K,2014A&A...565L...1P,2015ApJ...799...98M}. Finally, hot-Jupiter hosts have a smaller mean stellar birth radius ($4.60$ versus $5.69$ kpc) and a more negative corrected offset ($-0.76$ versus $-0.47$ kpc). Considered together with their higher mean $Z$, this may suggest a possible association between hot-Jupiter systems and metal-rich stellar populations formed in the inner Galactic disk.

\subsection{Statistical Sensitivity of Host--Non-host Abundance Differences}

An important limitation of the present analysis is that many of the corrected host--non-host offsets are modest compared with their uncertainties. This may result from a combination of intrinsically small differences, substantial star-to-star abundance scatter, and the limited number of detected planet hosts in individual planetary subsamples. In addition, our non-host sample consists of stars without detected Kepler planets rather than confirmed planet-free stars. Undetected planets caused by geometric transit probability or observational incompleteness may therefore exist in this sample. Such systems would make the nominal non-host population more similar to the detected host population, reducing the measured host--non-host offsets. Therefore, the modest differences found here should not be interpreted as evidence for identical underlying stellar populations, but rather as conservative estimates of the differences between stars with detected planets and stars without detected planets.

To evaluate the detectability of such abundance differences with the current sample size, we performed a Monte Carlo sensitivity analysis. For C, O, Mg, and Si, we selected the abundance bin showing the most prominent host--non-host difference among Earth-like planet hosts ($R_{\rm p}<2R_{\oplus}$). We fixed the observed host stars in each selected bin and generated mock non-host populations under a range of assumed intrinsic population differences, $\Delta[X/{\rm Fe}]$, between planet-hosting and non-hosting stars. The number of mock non-host stars was set to twice the number of selected hosts, following the approximate 1:2 ratio between detected planet hosts and non-host stars. The abundance scatter of each mock population was adopted from the corresponding full non-host stellar sample. For each assumed intrinsic difference, we performed 10,000 Monte Carlo realizations and calculated the fraction of realizations in which the host--non-host difference exceeds a $3\sigma$ significance level.

\begin{figure}
    \centering
    \includegraphics[width=\columnwidth]{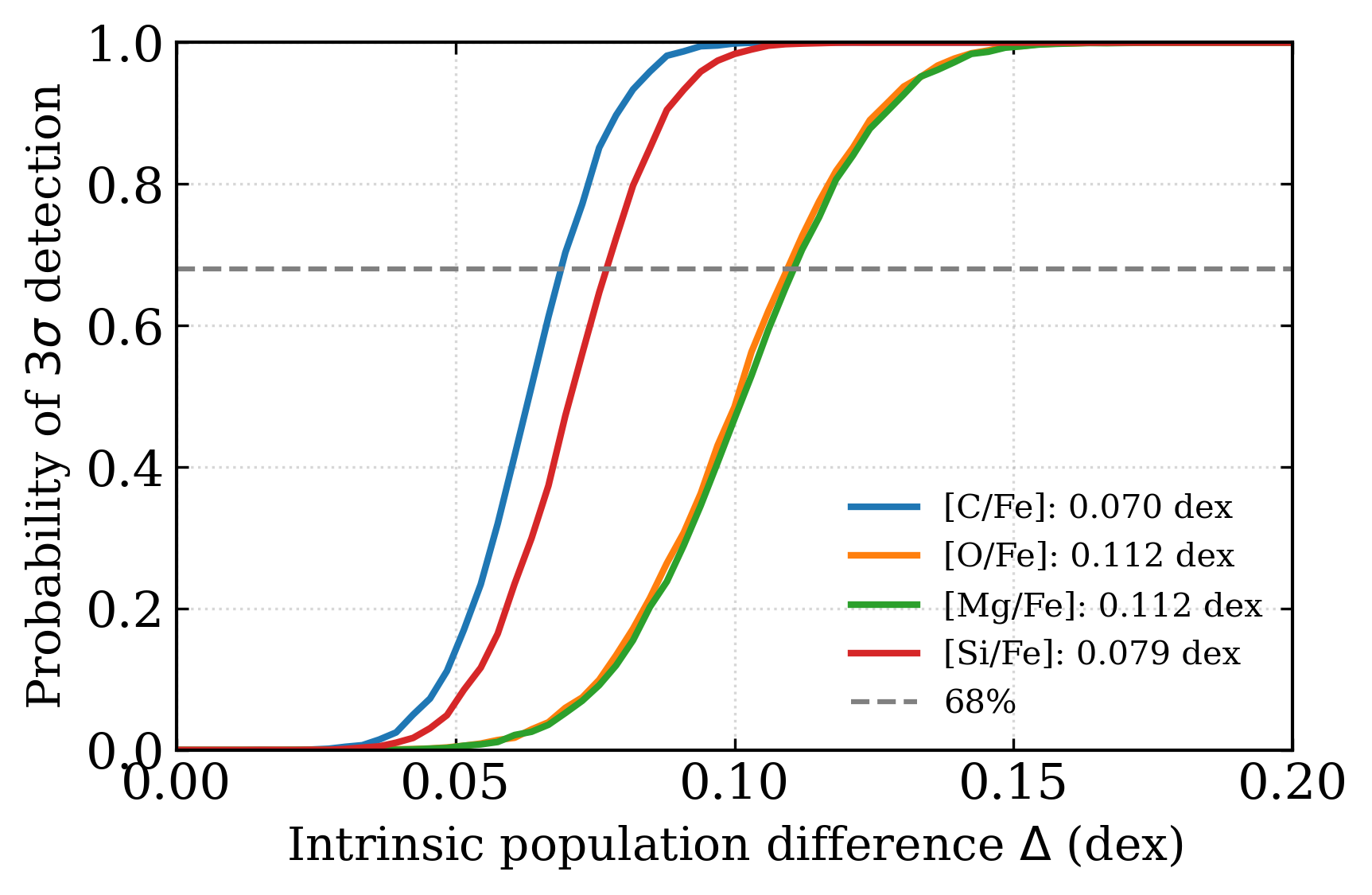}
    \caption{
    Monte Carlo estimate of the detectability of intrinsic host--non-host abundance differences.
    The curves show the probability of obtaining a $3\sigma$ detection as a function of the assumed intrinsic population difference for [C/Fe], [O/Fe], [Mg/Fe], and [Si/Fe].
    The horizontal dashed line indicates a 68\% detection probability, while the corresponding abundance differences required to reach this probability represent the approximate sensitivity limits of the current sample.
    }
    \label{fig:test}
\end{figure}

The resulting detection probabilities for [C/Fe], [O/Fe], [Mg/Fe], and [Si/Fe] are shown in Figure~\ref{fig:test}. We find that intrinsic population differences of $\Delta[{\rm C/Fe}] \approx 0.070$ dex, $\Delta[{\rm O/Fe}] \approx 0.112$ dex, $\Delta[{\rm Mg/Fe}] \approx 0.112$ dex, and $\Delta[{\rm Si/Fe}] \approx 0.079$ dex are required to achieve a 68\% probability of detecting a $3\sigma$ signal with the current sample sizes. These values should be interpreted as approximate sensitivity limits of our analysis rather than direct measurements of the intrinsic host--non-host abundance differences. The different sensitivity thresholds mainly reflect the combined effects of abundance scatter and the number of available planet hosts in each selected subsample. The results demonstrate that abundance differences below these levels may remain undetectable even if they are physically present. Therefore, the weak abundance trends identified in this work should be interpreted within the statistical sensitivity of the current sample, rather than as evidence against intrinsic chemical differences between planet-hosting and non-hosting stellar populations.

\section{Summary}
\label{sec:summary}

In this work, we investigated whether host stars differ from non-host stars of similar age and mass in their chemistry, activity, and birth radii, using a homogeneous Kepler--LAMOST--Gaia sample of 28,383 dwarfs and subgiants, including 629 host stars with 865 planets. We constructed an age--mass matched control sample for each host star and applied a fake-host baseline correction to remove distribution-induced matching biases. Our main results are summarized as follows.

\begin{enumerate}

\item We find that raw host--non-host differences can be strongly affected by the intrinsic distribution of the parameter being analyzed. Using [Fe/H] as an example, we show that much of the apparent raw metallicity trend is reproduced by the fake-host experiment using only non-host stars. After subtracting this baseline bias, the corrected [Fe/H] offsets are substantially reduced. This demonstrates the importance of correcting for distribution-induced matching biases before interpreting host--non-host differences.

\item For the overall metallicity $Z$, the full planet sample, Earth-like planet hosts, and sub-Neptune hosts show corrected offsets that are generally consistent with zero. The main exception is the giant-planet subsample, whose hosts tend to have slightly lower $Z$ than age--mass matched non-host stars. Converting the observed $\Delta Z$ values into equivalent $\Delta{\rm [Fe/H]}$ offsets suggests that this metallicity deficit is strongest at the metal-poor end, reaching $\sim0.03$ dex at ${\rm [Fe/H]}\approx-0.13$, and decreases toward higher metallicity. 

\item For elemental abundance ratios, the full planet sample shows no strong host–non-host differences in C, O, Mg, or Si. However, radius-dependent trends emerge after separating the sample by planet size. Earth-like and sub-Neptune hosts exhibit different abundance-dependent trends in O, Mg, and Si. At the high-abundance end, Earth-like hosts tend to be relatively O-rich but Mg- and Si-poor compared with their age–mass matched non-host stars, whereas sub-Neptune hosts tend to show the opposite tendency. Although these trends are modest, they suggest that Earth-like planets and sub-Neptunes may be associated with different chemical environments or different rocky building blocks.

\item For stellar magnetic activity, Earth-like hosts show no significant difference from matched non-host stars. In contrast, sub-Neptune hosts show a weak tendency toward negative corrected offsets in the relatively active regime, suggesting that they may be slightly less magnetically active than age–mass matched non-host stars. This lower activity, if confirmed with larger samples, could correspond to a weaker high-energy irradiation environment and may help sub-Neptunes preserve primordial gaseous envelopes.

\item For stellar birth radius, Earth-like hosts are generally consistent with their matched non-host stars, whereas sub-Neptune hosts show systematically negative $\Delta R_{\rm b}$ over $R_{\rm b}\sim4$--$7$ kpc. This indicates that sub-Neptune hosts were, on average, born closer to the Galactic center than otherwise similar non-host stars. Together with their Mg/Si abundance behavior and modest activity trend, this result may point to formation environments that are more favorable for building and retaining volatile-rich envelopes.

\item Hot- and longer-period Jupiter hosts show broadly similar elemental-abundance offsets, while tentative differences emerge in their overall metallicity, chromospheric activity, and stellar birth radius. Hot-Jupiter hosts tend to be more metal-rich and active and to have formed at smaller Galactic radii, suggesting a possible association between hot-Jupiter systems and metal-rich inner-disk stellar populations.

\end{enumerate}

Overall, our results show that host--non-host differences are generally weak when all planets are considered together, indicating that planet-hosting stars are, in most respects, remarkably similar to non-host stars of comparable age and mass. Significant differences become apparent only after separating planets by radius. In particular, Earth-like planets and sub-Neptunes exhibit different trends in chemistry, activity, and birth radius, suggesting that these two populations may reflect different formation environments and evolutionary histories. Our comparison of giant-planet systems also suggests that hot-Jupiter hosts may be more metal-rich and active and may have formed at smaller Galactic radii than hosts of longer-period Jupiters, although these differences remain tentative. Furthermore, Monte Carlo sensitivity tests for representative strong abundance trends in Earth-like hosts indicate that intrinsic abundance differences of $\sim$0.07--$0.11$ dex are required to achieve a $3\sigma$ detection with the current sample sizes. Therefore, the weak abundance trends identified in this work should be interpreted in the context of the current statistical sensitivity, and smaller intrinsic host--non-host chemical differences may remain unresolved because of intrinsic stellar abundance scatter and limited sample sizes.

The statistical significance of some of these trends is limited by the current sample size, the intrinsic star-to-star scatter, and especially the small number of giant-planet systems. In addition, the nominal non-host sample consists of stars without detected Kepler planets and may therefore contain systems with undetected planets. If the measured host--non-host differences are real, the inclusion of such systems would dilute both their amplitudes and statistical significance, implying that the observed offsets may provide conservative estimates of the underlying differences. Larger and more homogeneous planet-host samples from future surveys, combined with ongoing spectroscopic programs and forward modeling of radius-dependent planet occurrence rates and star-by-star detection completeness, will be essential for testing these trends and clarifying the roles of stellar chemistry, magnetic activity, and Galactic environment in shaping planetary systems.

\begin{acknowledgments}
This work is based on data acquired through the Guoshoujing Telescope. Guoshoujing Telescope (the Large Sky Area Multi-Object Fiber Spectroscopic Telescope; LAMOST) is a National Major Scientific Project built by the Chinese Academy of Sciences. Funding for the project has been provided by the National Development and Reform Commission. LAMOST is operated and managed by the National Astronomical Observatories, Chinese Academy of Sciences. This work has made use of data from the European Space Agency (ESA) mission Gaia (\url{https://www.cosmos.esa.int/gaia}), processed by the Gaia Data Processing and Analysis Consortium (DPAC, \url{https://www.cosmos.esa.int/web/gaia/dpac/consortium}). Funding for the DPAC has been provided by national institutions, in particular the institutions participating in the Gaia Multilateral Agreement. We acknowledge the entire Kepler team and everyone involved in the Kepler mission. Funding for the Kepler mission is provided by NASA's Science Mission Directorate.
This work was supported by the National Natural Science Foundation of China (Grants No.12403037, Grants No.12503025).
\end{acknowledgments}

\clearpage
\bibliography{Biblio}{}

\begin{thebibliography}{}
\expandafter\ifx\csname natexlab\endcsname\relax\def\natexlab#1{#1}\fi
\providecommand{\url}[1]{\href{#1}{#1}}
\providecommand{\dodoi}[1]{doi:~\href{http://doi.org/#1}{\nolinkurl{#1}}}
\providecommand{\doeprint}[1]{\href{http://ascl.net/#1}{\nolinkurl{http://ascl.net/#1}}}
\providecommand{\doarXiv}[1]{\href{https://arxiv.org/abs/#1}{\nolinkurl{https://arxiv.org/abs/#1}}}

\bibitem[{R.~L. {Akeson} {et~al.}(2013){Akeson}, {Chen}, {Ciardi}, {Crane},
  {Good}, {Harbut}, {Jackson}, {Kane}, {Laity}, {Leifer}, {Lynn}, {McElroy},
  {Papin}, {Plavchan}, {Ram{\'\i}rez}, {Rey}, {von Braun}, {Wittman},
  {Abajian}, {Ali}, {Beichman}, {Beekley}, {Berriman}, {Berukoff}, {Bryden},
  {Chan}, {Groom}, {Lau}, {Payne}, {Regelson}, {Saucedo}, {Schmitz},
  {Stauffer}, {Wyatt}, \& {Zhang}}]{2013PASP..125..989A}
{Akeson}, R.~L., {Chen}, X., {Ciardi}, D., {et~al.} 2013, \bibinfo{title}{{The
  NASA Exoplanet Archive: Data and Tools for Exoplanet Research},} \pasp, 125,
  989, \dodoi{10.1086/672273}

\bibitem[{S.~M. {Andrews} {et~al.}(2013){Andrews}, {Rosenfeld}, {Kraus}, \&
  {Wilner}}]{2013ApJ...771..129A}
{Andrews}, S.~M., {Rosenfeld}, K.~A., {Kraus}, A.~L., \& {Wilner}, D.~J. 2013,
  \bibinfo{title}{{The Mass Dependence between Protoplanetary Disks and their
  Stellar Hosts},} \apj, 771, 129, \dodoi{10.1088/0004-637X/771/2/129}

\bibitem[{S. {Basu} {et~al.}(2010){Basu}, {Chaplin}, \&
  {Elsworth}}]{2010ApJ...710.1596B}
{Basu}, S., {Chaplin}, W.~J., \& {Elsworth}, Y. 2010,
  \bibinfo{title}{{Determination of Stellar Radii from Asteroseismic Data},}
  \apj, 710, 1596, \dodoi{10.1088/0004-637X/710/2/1596}

\bibitem[{T.~A. {Berger} {et~al.}(2020){Berger}, {Huber}, {van Saders},
  {Gaidos}, {Tayar}, \& {Kraus}}]{2020AJ....159..280B}
{Berger}, T.~A., {Huber}, D., {van Saders}, J.~L., {et~al.} 2020,
  \bibinfo{title}{{The Gaia-Kepler Stellar Properties Catalog. I. Homogeneous
  Fundamental Properties for 186,301 Kepler Stars},} \aj, 159, 280,
  \dodoi{10.3847/1538-3881/159/6/280}

\bibitem[{K.~M. {Boley} {et~al.}(2024){Boley}, {Christiansen}, {Zink},
  {Hardegree-Ullman}, {Lee}, {Hopkins}, {Wang}, {Fernandes}, {Bergsten}, \&
  {Bhure}}]{2024AJ....168..128B}
{Boley}, K.~M., {Christiansen}, J.~L., {Zink}, J., {et~al.} 2024,
  \bibinfo{title}{{The First Evidence of a Host Star Metallicity Cutoff in the
  Formation of Super-Earth Planets},} \aj, 168, 128,
  \dodoi{10.3847/1538-3881/ad6570}

\bibitem[{W.~J. {Borucki} {et~al.}(2010){Borucki}, {Koch}, {Basri}, {Batalha},
  {Brown}, {Caldwell}, {Caldwell}, {Christensen-Dalsgaard}, {Cochran},
  {DeVore}, {Dunham}, {Dupree}, {Gautier}, {Geary}, {Gilliland}, {Gould},
  {Howell}, {Jenkins}, {Kondo}, {Latham}, {Marcy}, {Meibom}, {Kjeldsen},
  {Lissauer}, {Monet}, {Morrison}, {Sasselov}, {Tarter}, {Boss}, {Brownlee},
  {Owen}, {Buzasi}, {Charbonneau}, {Doyle}, {Fortney}, {Ford}, {Holman},
  {Seager}, {Steffen}, {Welsh}, {Rowe}, {Anderson}, {Buchhave}, {Ciardi},
  {Walkowicz}, {Sherry}, {Horch}, {Isaacson}, {Everett}, {Fischer}, {Torres},
  {Johnson}, {Endl}, {MacQueen}, {Bryson}, {Dotson}, {Haas}, {Kolodziejczak},
  {Van Cleve}, {Chandrasekaran}, {Twicken}, {Quintana}, {Clarke}, {Allen},
  {Li}, {Wu}, {Tenenbaum}, {Verner}, {Bruhweiler}, {Barnes}, \&
  {Prsa}}]{2010Sci...327..977B}
{Borucki}, W.~J., {Koch}, D., {Basri}, G., {et~al.} 2010,
  \bibinfo{title}{{Kepler Planet-Detection Mission: Introduction and First
  Results},} Science, 327, 977, \dodoi{10.1126/science.1185402}

\bibitem[{J.~M. {Brewer} \& D.~A. {Fischer}(2016){Brewer} \&
  {Fischer}}]{2016ApJ...831...20B}
{Brewer}, J.~M., \& {Fischer}, D.~A. 2016, \bibinfo{title}{{C/O and Mg/Si
  Ratios of Stars in the Solar Neighborhood},} \apj, 831, 20,
  \dodoi{10.3847/0004-637X/831/1/20}

\bibitem[{H. {Bruntt} {et~al.}(2012){Bruntt}, {Basu}, {Smalley}, {Chaplin},
  {Verner}, {Bedding}, {Catala}, {Gazzano}, {Molenda-{\.Z}akowicz}, {Thygesen},
  {Uytterhoeven}, {Hekker}, {Huber}, {Karoff}, {Mathur}, {Mosser},
  {Appourchaux}, {Campante}, {Elsworth}, {Garc{\'\i}a}, {Handberg}, {Metcalfe},
  {Quirion}, {R{\'e}gulo}, {Roxburgh}, {Stello}, {Christensen-Dalsgaard},
  {Kawaler}, {Kjeldsen}, {Morris}, {Quintana}, \&
  {Sanderfer}}]{2012MNRAS.423..122B}
{Bruntt}, H., {Basu}, S., {Smalley}, B., {et~al.} 2012,
  \bibinfo{title}{{Accurate fundamental parameters and detailed abundance
  patterns from spectroscopy of 93 solar-type Kepler targets},} \mnras, 423,
  122, \dodoi{10.1111/j.1365-2966.2012.20686.x}

\bibitem[{S. {Bryson} {et~al.}(2020){Bryson}, {Coughlin}, {Batalha}, {Berger},
  {Huber}, {Burke}, {Dotson}, \& {Mullally}}]{2020AJ....159..279B}
{Bryson}, S., {Coughlin}, J., {Batalha}, N.~M., {et~al.} 2020,
  \bibinfo{title}{{A Probabilistic Approach to Kepler Completeness and
  Reliability for Exoplanet Occurrence Rates},} \aj, 159, 279,
  \dodoi{10.3847/1538-3881/ab8a30}

\bibitem[{L.~A. {Buchhave} {et~al.}(2014){Buchhave}, {Bizzarro}, {Latham},
  {Sasselov}, {Cochran}, {Endl}, {Isaacson}, {Juncher}, \&
  {Marcy}}]{2014Natur.509..593B}
{Buchhave}, L.~A., {Bizzarro}, M., {Latham}, D.~W., {et~al.} 2014,
  \bibinfo{title}{{Three regimes of extrasolar planet radius inferred from host
  star metallicities},} \nat, 509, 593, \dodoi{10.1038/nature13254}

\bibitem[{J.~C. {Carter-Bond} {et~al.}(2012){Carter-Bond}, {O'Brien}, {Delgado
  Mena}, {Israelian}, {Santos}, \& {Gonz{\'a}lez
  Hern{\'a}ndez}}]{2012ApJ...747L...2C}
{Carter-Bond}, J.~C., {O'Brien}, D.~P., {Delgado Mena}, E., {et~al.} 2012,
  \bibinfo{title}{{Low Mg/Si Planetary Host Stars and Their Mg-depleted
  Terrestrial Planets},} \apjl, 747, L2, \dodoi{10.1088/2041-8205/747/1/L2}

\bibitem[{C. {Catala} {et~al.}(2007){Catala}, {Donati}, {Shkolnik},
  {Bohlender}, \& {Alecian}}]{2007MNRAS.374L..42C}
{Catala}, C., {Donati}, J.-F., {Shkolnik}, E., {Bohlender}, D., \& {Alecian},
  E. 2007, \bibinfo{title}{{The magnetic field of the planet-hosting star
  {\ensuremath{\tau}} Bootis},} \mnras, 374, L42,
  \dodoi{10.1111/j.1745-3933.2006.00261.x}

\bibitem[{D.-C. {Chen} {et~al.}(2026){Chen}, {Xie}, {Zhou}, {Dai}, {Ma},
  {Wang}, \& {Liu}}]{2026NatAs..10...92C}
{Chen}, D.-C., {Xie}, J.-W., {Zhou}, J.-L., {et~al.} 2026, \bibinfo{title}{{The
  origin and tidal evolution of hot Jupiters constrained by a broken
  age-frequency relation},} Nature Astronomy, 10, 92,
  \dodoi{10.1038/s41550-025-02693-6}

\bibitem[{D.-C. {Chen} {et~al.}(2021){Chen}, {Yang}, {Xie}, {Zhou}, {Dong},
  {Zheng}, {Zhang}, {Liu}, {Wang}, {Xiang}, {Zong}, {Huang}, \&
  {Luo}}]{2021AJ....162..100C}
{Chen}, D.-C., {Yang}, J.-Y., {Xie}, J.-W., {et~al.} 2021,
  \bibinfo{title}{{Planets Across Space and Time (PAST). II. Catalog and
  Analyses of the LAMOST-Gaia-Kepler Stellar Kinematic Properties},} \aj, 162,
  100, \dodoi{10.3847/1538-3881/ac0f08}

\bibitem[{H. {Chen} \& L.~A. {Rogers}(2016){Chen} \&
  {Rogers}}]{2016ApJ...831..180C}
{Chen}, H., \& {Rogers}, L.~A. 2016, \bibinfo{title}{{Evolutionary Analysis of
  Gaseous Sub-Neptune-mass Planets with MESA},} \apj, 831, 180,
  \dodoi{10.3847/0004-637X/831/2/180}

\bibitem[{X. {Chen} {et~al.}(2026){Chen}, {Sun}, {Lu}, {Lu}, \&
  {Ye}}]{2026ApJ..1001..176C}
{Chen}, X., {Sun}, T., {Lu}, Y.~L., {Lu}, Z., \& {Ye}, L. 2026,
  \bibinfo{title}{{Insights into the Exoplanet Radius Valley from Host-star
  Ages, Activity, Chemistry, and Birth Radii},} \apj, 1001, 176,
  \dodoi{10.3847/1538-4357/ae5633}

\bibitem[{X. {Chen} {et~al.}(2025){Chen}, {Sun}, \& {Ye}}]{2025ApJ...995...33C}
{Chen}, X., {Sun}, T., \& {Ye}, L. 2025, \bibinfo{title}{{Insights into Planet
  Formation from the Ages, Masses, and Elemental Abundances of Host Stars},}
  \apj, 995, 33, \dodoi{10.3847/1538-4357/ae12a2}

\bibitem[{X.-Q. {Cui} {et~al.}(2012){Cui}, {Zhao}, {Chu}, {Li}, {Li}, {Zhang},
  {Su}, {Yao}, {Wang}, {Xing}, {Li}, {Zhu}, {Wang}, {Gu}, {Luo}, {Xu}, {Zhang},
  {Liu}, {Zhang}, {Yang}, {Cao}, {Chen}, {Chen}, {Chen}, {Chen}, {Chu}, {Feng},
  {Gong}, {Hou}, {Hu}, {Hu}, {Hu}, {Jia}, {Jiang}, {Jiang}, {Jiang}, {Jin},
  {Li}, {Li}, {Li}, {Liu}, {Liu}, {Lu}, {Mao}, {Men}, {Qi}, {Qi}, {Shi},
  {Tang}, {Tao}, {Wang}, {Wang}, {Wang}, {Wang}, {Wang}, {Wang}, {Wang},
  {Wang}, {Wang}, {Wang}, {Wang}, {Wang}, {Xu}, {Xu}, {Yang}, {Yu}, {Yuan},
  {Yuan}, {Zhai}, {Zhang}, {Zhang}, {Zhang}, {Zhao}, {Zhou}, {Zhou}, {Zhu}, \&
  {Zou}}]{2012RAA....12.1197C}
{Cui}, X.-Q., {Zhao}, Y.-H., {Chu}, Y.-Q., {et~al.} 2012, \bibinfo{title}{{The
  Large Sky Area Multi-Object Fiber Spectroscopic Telescope (LAMOST)},}
  Research in Astronomy and Astrophysics, 12, 1197,
  \dodoi{10.1088/1674-4527/12/9/003}

\bibitem[{T.~J. David {et~al.}(2021)David, Contardo, Sandoval, Angus, Lu,
  Bedell, Curtis, Foreman-Mackey, Fulton, Grunblatt, \& Petigura}]{David_2021}
David, T.~J., Contardo, G., Sandoval, A., {et~al.} 2021,
  \bibinfo{title}{Evolution of the Exoplanet Size Distribution: Forming Large
  Super-Earths Over Billions of Years,} The Astronomical Journal, 161, 265,
  \dodoi{10.3847/1538-3881/abf439}

\bibitem[{R.~I. {Dawson} \& R.~A. {Murray-Clay}(2013){Dawson} \&
  {Murray-Clay}}]{2013ApJ...767L..24D}
{Dawson}, R.~I., \& {Murray-Clay}, R.~A. 2013, \bibinfo{title}{{Giant Planets
  Orbiting Metal-rich Stars Show Signatures of Planet-Planet Interactions},}
  \apjl, 767, L24, \dodoi{10.1088/2041-8205/767/2/L24}

\bibitem[{G.~M. {De Silva} {et~al.}(2015){De Silva}, {Freeman},
  {Bland-Hawthorn}, {Martell}, {de Boer}, {Asplund}, {Keller}, {Sharma},
  {Zucker}, {Zwitter}, {Anguiano}, {Bacigalupo}, {Bayliss}, {Beavis},
  {Bergemann}, {Campbell}, {Cannon}, {Carollo}, {Casagrande}, {Casey}, {Da
  Costa}, {D'Orazi}, {Dotter}, {Duong}, {Heger}, {Ireland}, {Kafle}, {Kos},
  {Lattanzio}, {Lewis}, {Lin}, {Lind}, {Munari}, {Nataf}, {O'Toole}, {Parker},
  {Reid}, {Schlesinger}, {Sheinis}, {Simpson}, {Stello}, {Ting}, {Traven},
  {Watson}, {Wittenmyer}, {Yong}, \& {{\v{Z}}erjal}}]{2015MNRAS.449.2604D}
{De Silva}, G.~M., {Freeman}, K.~C., {Bland-Hawthorn}, J., {et~al.} 2015,
  \bibinfo{title}{{The GALAH survey: scientific motivation},} \mnras, 449,
  2604, \dodoi{10.1093/mnras/stv327}

\bibitem[{P. {Demarque} {et~al.}(2004){Demarque}, {Woo}, {Kim}, \&
  {Yi}}]{2004ApJS..155..667D}
{Demarque}, P., {Woo}, J.-H., {Kim}, Y.-C., \& {Yi}, S.~K. 2004,
  \bibinfo{title}{{Y$^{2}$ Isochrones with an Improved Core Overshoot
  Treatment},} \apjs, 155, 667, \dodoi{10.1086/424966}

\bibitem[{S. {Dong} {et~al.}(2014){Dong}, {Zheng}, {Zhu}, {De Cat}, {Fu},
  {Yang}, {Zhang}, {Jin}, \& {Zhang}}]{2014ApJ...789L...3D}
{Dong}, S., {Zheng}, Z., {Zhu}, Z., {et~al.} 2014, \bibinfo{title}{{On the
  Metallicities of Kepler Stars},} \apjl, 789, L3,
  \dodoi{10.1088/2041-8205/789/1/L3}

\bibitem[{M.~E. {Everett} {et~al.}(2013){Everett}, {Howell}, {Silva}, \&
  {Szkody}}]{2013ApJ...771..107E}
{Everett}, M.~E., {Howell}, S.~B., {Silva}, D.~R., \& {Szkody}, P. 2013,
  \bibinfo{title}{{Spectroscopy of Faint Kepler Mission Exoplanet Candidate
  Host Stars},} \apj, 771, 107, \dodoi{10.1088/0004-637X/771/2/107}

\bibitem[{D.~K. {Feuillet} {et~al.}(2019){Feuillet}, {Frankel}, {Lind},
  {Frinchaboy}, {Garc{\'\i}a-Hern{\'a}ndez}, {Lane}, {Nitschelm}, \&
  {Roman-Lopes}}]{2019MNRAS.489.1742F}
{Feuillet}, D.~K., {Frankel}, N., {Lind}, K., {et~al.} 2019,
  \bibinfo{title}{{Spatial variations in the Milky Way disc metallicity-age
  relation},} \mnras, 489, 1742, \dodoi{10.1093/mnras/stz2221}

\bibitem[{D.~K. {Feuillet} {et~al.}(2018){Feuillet}, {Bovy}, {Holtzman},
  {Weinberg}, {Garc{\'\i}a-Hern{\'a}ndez}, {Hearty}, {Majewski}, {Roman-Lopes},
  {Rybizki}, \& {Zamora}}]{2018MNRAS.477.2326F}
{Feuillet}, D.~K., {Bovy}, J., {Holtzman}, J., {et~al.} 2018,
  \bibinfo{title}{{Age-resolved chemistry of red giants in the solar
  neighbourhood},} \mnras, 477, 2326, \dodoi{10.1093/mnras/sty779}

\bibitem[{D.~A. {Fischer} \& J. {Valenti}(2005){Fischer} \&
  {Valenti}}]{2005ApJ...622.1102F}
{Fischer}, D.~A., \& {Valenti}, J. 2005, \bibinfo{title}{{The
  Planet-Metallicity Correlation},} \apj, 622, 1102, \dodoi{10.1086/428383}

\bibitem[{S.~W. {Fleming} {et~al.}(2015){Fleming}, {Mahadevan}, {Deshpande},
  {Bender}, {Terrien}, {Marchwinski}, {Wang}, {Roy}, {Stassun}, {Allende
  Prieto}, {Cunha}, {Smith}, {Agol}, {Ak}, {Bastien}, {Bizyaev}, {Crepp},
  {Ford}, {Frinchaboy}, {Garc{\'\i}a-Hern{\'a}ndez}, {Garc{\'\i}a P{\'e}rez},
  {Gaudi}, {Ge}, {Hearty}, {Ma}, {Majewski}, {M{\'e}sz{\'a}ros}, {Nidever},
  {Pan}, {Pepper}, {Pinsonneault}, {Schiavon}, {Schneider}, {Wilson}, {Zamora},
  \& {Zasowski}}]{2015AJ....149..143F}
{Fleming}, S.~W., {Mahadevan}, S., {Deshpande}, R., {et~al.} 2015,
  \bibinfo{title}{{The APOGEE Spectroscopic Survey of Kepler Planet Hosts:
  Feasibility, Efficiency, and First Results},} \aj, 149, 143,
  \dodoi{10.1088/0004-6256/149/4/143}

\bibitem[{B.~J. {Fulton} {et~al.}(2017){Fulton}, {Petigura}, {Howard},
  {Isaacson}, {Marcy}, {Cargile}, {Hebb}, {Weiss}, {Johnson}, {Morton},
  {Sinukoff}, {Crossfield}, \& {Hirsch}}]{2017AJ....154..109F}
{Fulton}, B.~J., {Petigura}, E.~A., {Howard}, A.~W., {et~al.} 2017,
  \bibinfo{title}{{The California-Kepler Survey. III. A Gap in the Radius
  Distribution of Small Planets},} \aj, 154, 109,
  \dodoi{10.3847/1538-3881/aa80eb}

\bibitem[{ {Gaia Collaboration} {et~al.}(2023){Gaia Collaboration},
  {Vallenari}, {Brown}, {Prusti}, {de Bruijne}, {Arenou}, {Babusiaux},
  {Biermann}, {Creevey}, {Ducourant}, {Evans}, {Eyer}, {Guerra}, {Hutton},
  {Jordi}, {Klioner}, {Lammers}, {Lindegren}, {Luri}, {Mignard}, {Panem},
  {Pourbaix}, {Randich}, {Sartoretti}, {Soubiran}, {Tanga}, {Walton},
  {Bailer-Jones}, {Bastian}, {Drimmel}, {Jansen}, {Katz}, {Lattanzi}, {van
  Leeuwen}, {Bakker}, {Cacciari}, {Casta{\~n}eda}, {De Angeli}, {Fabricius},
  {Fouesneau}, {Fr{\'e}mat}, {Galluccio}, {Guerrier}, {Heiter}, {Masana},
  {Messineo}, {Mowlavi}, {Nicolas}, {Nienartowicz}, {Pailler}, {Panuzzo},
  {Riclet}, {Roux}, {Seabroke}, {Sordo}, {Th{\'e}venin}, {Gracia-Abril},
  {Portell}, {Teyssier}, {Altmann}, {Andrae}, {Audard}, {Bellas-Velidis},
  {Benson}, {Berthier}, {Blomme}, {Burgess}, {Busonero}, {Busso},
  {C{\'a}novas}, {Carry}, {Cellino}, {Cheek}, {Clementini}, {Damerdji},
  {Davidson}, {de Teodoro}, {Nu{\~n}ez Campos}, {Delchambre}, {Dell'Oro},
  {Esquej}, {Fern{\'a}ndez-Hern{\'a}ndez}, {Fraile}, {Garabato},
  {Garc{\'\i}a-Lario}, {Gosset}, {Haigron}, {Halbwachs}, {Hambly}, {Harrison},
  {Hern{\'a}ndez}, {Hestroffer}, {Hodgkin}, {Holl}, {Jan{\ss}en}, {Jevardat de
  Fombelle}, {Jordan}, {Krone-Martins}, {Lanzafame}, {L{\"o}ffler}, {Marchal},
  {Marrese}, {Moitinho}, {Muinonen}, {Osborne}, {Pancino}, {Pauwels},
  {Recio-Blanco}, {Reyl{\'e}}, {Riello}, {Rimoldini}, {Roegiers}, {Rybizki},
  {Sarro}, {Siopis}, {Smith}, {Sozzetti}, {Utrilla}, {van Leeuwen}, {Abbas},
  {{\'A}brah{\'a}m}, {Abreu Aramburu}, {Aerts}, {Aguado}, {Ajaj},
  {Aldea-Montero}, {Altavilla}, {{\'A}lvarez}, {Alves}, {Anders}, {Anderson},
  {Anglada Varela}, {Antoja}, {Baines}, {Baker}, {Balaguer-N{\'u}{\~n}ez},
  {Balbinot}, {Balog}, {Barache}, {Barbato}, {Barros}, {Barstow},
  {Bartolom{\'e}}, {Bassilana}, {Bauchet}, {Becciani}, {Bellazzini},
  {Berihuete}, {Bernet}, {Bertone}, {Bianchi}, {Binnenfeld}, {Blanco-Cuaresma},
  {Blazere}, {Boch}, {Bombrun}, {Bossini}, {Bouquillon}, {Bragaglia},
  {Bramante}, {Breedt}, {Bressan}, {Brouillet}, {Brugaletta}, {Bucciarelli},
  {Burlacu}, {Butkevich}, {Buzzi}, {Caffau}, {Cancelliere}, {Cantat-Gaudin},
  {Carballo}, {Carlucci}, {Carnerero}, {Carrasco}, {Casamiquela}, {Castellani},
  {Castro-Ginard}, {Chaoul}, {Charlot}, {Chemin}, {Chiaramida}, {Chiavassa},
  {Chornay}, {Comoretto}, {Contursi}, {Cooper}, {Cornez}, {Cowell}, {Crifo},
  {Cropper}, {Crosta}, {Crowley}, {Dafonte}, {Dapergolas}, {David}, {David},
  {de Laverny}, {De Luise}, \& {De March}}]{2023gaia}
{Gaia Collaboration}, {Vallenari}, A., {Brown}, A.~G.~A., {et~al.} 2023,
  \bibinfo{title}{{Gaia Data Release 3. Summary of the content and survey
  properties},} \aap, 674, A1, \dodoi{10.1051/0004-6361/202243940}

\bibitem[{H.-P. {Gail} \& E. {Sedlmayr}(2013){Gail} \&
  {Sedlmayr}}]{2013pccd.book.....G}
{Gail}, H.-P., \& {Sedlmayr}, E. 2013, {Physics and Chemistry of Circumstellar
  Dust Shells}

\bibitem[{L. {Ghezzi} {et~al.}(2026){Ghezzi}, {Costa-Almeida}, {Loaiza-Tacuri},
  \& {Cunha}}]{2026ApJ...998..301G}
{Ghezzi}, L., {Costa-Almeida}, E., {Loaiza-Tacuri}, V., \& {Cunha}, K. 2026,
  \bibinfo{title}{{A Comprehensive Study of the Relations between the
  Properties of Planetary Systems and the Chemical Compositions of Their Host
  Stars},} \apj, 998, 301, \dodoi{10.3847/1538-4357/ae317d}

\bibitem[{S. {Ginzburg} {et~al.}(2018){Ginzburg}, {Schlichting}, \&
  {Sari}}]{2018MNRAS.476..759G}
{Ginzburg}, S., {Schlichting}, H.~E., \& {Sari}, R. 2018,
  \bibinfo{title}{{Core-powered mass-loss and the radius distribution of small
  exoplanets},} \mnras, 476, 759, \dodoi{10.1093/mnras/sty290}

\bibitem[{A. {Gupta} \& H.~E. {Schlichting}(2019){Gupta} \&
  {Schlichting}}]{2019MNRAS.487...24G}
{Gupta}, A., \& {Schlichting}, H.~E. 2019, \bibinfo{title}{{Sculpting the
  valley in the radius distribution of small exoplanets as a by-product of
  planet formation: the core-powered mass-loss mechanism},} \mnras, 487, 24,
  \dodoi{10.1093/mnras/stz1230}

\bibitem[{A. {Gupta} \& H.~E. {Schlichting}(2020){Gupta} \&
  {Schlichting}}]{2020MNRAS.493..792G}
{Gupta}, A., \& {Schlichting}, H.~E. 2020, \bibinfo{title}{{Signatures of the
  core-powered mass-loss mechanism in the exoplanet population: dependence on
  stellar properties and observational predictions},} \mnras, 493, 792,
  \dodoi{10.1093/mnras/staa315}

\bibitem[{M. {Haywood} {et~al.}(2013){Haywood}, {Di Matteo}, {Lehnert}, {Katz},
  \& {G{\'o}mez}}]{2013A&A...560A.109H}
{Haywood}, M., {Di Matteo}, P., {Lehnert}, M.~D., {Katz}, D., \& {G{\'o}mez},
  A. 2013, \bibinfo{title}{{The age structure of stellar populations in the
  solar vicinity. Clues of a two-phase formation history of the Milky Way
  disk},} \aap, 560, A109, \dodoi{10.1051/0004-6361/201321397}

\bibitem[{S. {Jin} {et~al.}(2014){Jin}, {Mordasini}, {Parmentier}, {van
  Boekel}, {Henning}, \& {Ji}}]{2014ApJ...795...65J}
{Jin}, S., {Mordasini}, C., {Parmentier}, V., {et~al.} 2014,
  \bibinfo{title}{{Planetary Population Synthesis Coupled with Atmospheric
  Escape: A Statistical View of Evaporation},} \apj, 795, 65,
  \dodoi{10.1088/0004-637X/795/1/65}

\bibitem[{J.~A. Johnson {et~al.}(2010)Johnson, Aller, Howard, \&
  Crepp}]{Johnson_2010}
Johnson, J.~A., Aller, K.~M., Howard, A.~W., \& Crepp, J.~R. 2010,
  \bibinfo{title}{Giant Planet Occurrence in the Stellar Mass-Metallicity
  Plane,} Publications of the Astronomical Society of the Pacific, 122, 905,
  \dodoi{10.1086/655775}

\bibitem[{J.~A. {Johnson} {et~al.}(2010){Johnson}, {Aller}, {Howard}, \&
  {Crepp}}]{2010PASP..122..905J}
{Johnson}, J.~A., {Aller}, K.~M., {Howard}, A.~W., \& {Crepp}, J.~R. 2010,
  \bibinfo{title}{{Giant Planet Occurrence in the Stellar Mass-Metallicity
  Plane},} \pasp, 122, 905, \dodoi{10.1086/655775}

\bibitem[{T. {Kallinger} {et~al.}(2010){Kallinger}, {Mosser}, {Hekker},
  {Huber}, {Stello}, {Mathur}, {Basu}, {Bedding}, {Chaplin}, {De Ridder},
  {Elsworth}, {Frandsen}, {Garc{\'\i}a}, {Gruberbauer}, {Matthews}, {Borucki},
  {Bruntt}, {Christensen-Dalsgaard}, {Gilliland}, {Kjeldsen}, \&
  {Koch}}]{2010A&A...522A...1K}
{Kallinger}, T., {Mosser}, B., {Hekker}, S., {et~al.} 2010,
  \bibinfo{title}{{Asteroseismology of red giants from the first four months of
  Kepler data: Fundamental stellar parameters},} \aap, 522, A1,
  \dodoi{10.1051/0004-6361/201015263}

\bibitem[{G. {Kov{\'a}cs} {et~al.}(2014){Kov{\'a}cs}, {Hartman}, {Bakos},
  {Quinn}, {Penev}, {Latham}, {Bhatti}, {Csubry}, \& {de
  Val-Borro}}]{2014MNRAS.442.2081K}
{Kov{\'a}cs}, G., {Hartman}, J.~D., {Bakos}, G.~{\'A}., {et~al.} 2014,
  \bibinfo{title}{{Stellar rotational periods in the planet hosting open
  cluster Praesepe},} \mnras, 442, 2081, \dodoi{10.1093/mnras/stu946}

\bibitem[{M.~J. {Kuchner} \& S. {Seager}(2005){Kuchner} \&
  {Seager}}]{2005astro.ph..4214K}
{Kuchner}, M.~J., \& {Seager}, S. 2005, \bibinfo{title}{{Extrasolar Carbon
  Planets},} arXiv e-prints, astro, \dodoi{10.48550/arXiv.astro-ph/0504214}

\bibitem[{L. {Lindegren} {et~al.}(2021){Lindegren}, {Bastian}, {Biermann},
  {Bombrun}, {de Torres}, {Gerlach}, {Geyer}, {Hern{\'a}ndez}, {Hilger},
  {Hobbs}, {Klioner}, {Lammers}, {McMillan}, {Ramos-Lerate},
  {Steidelm{\"u}ller}, {Stephenson}, \& {van Leeuwen}}]{2021A&A...649A...4L}
{Lindegren}, L., {Bastian}, U., {Biermann}, M., {et~al.} 2021,
  \bibinfo{title}{{Gaia Early Data Release 3. Parallax bias versus magnitude,
  colour, and position},} \aap, 649, A4, \dodoi{10.1051/0004-6361/202039653}

\bibitem[{E.~D. {Lopez} \& J.~J. {Fortney}(2013){Lopez} \&
  {Fortney}}]{2013ApJ...776....2L}
{Lopez}, E.~D., \& {Fortney}, J.~J. 2013, \bibinfo{title}{{The Role of Core
  Mass in Controlling Evaporation: The Kepler Radius Distribution and the
  Kepler-36 Density Dichotomy},} \apj, 776, 2,
  \dodoi{10.1088/0004-637X/776/1/2}

\bibitem[{Y.~L. {Lu} {et~al.}(2022){Lu}, {Ness}, {Buck}, {Zinn}, \&
  {Johnston}}]{2022MNRAS.512.2890L}
{Lu}, Y.~L., {Ness}, M.~K., {Buck}, T., {Zinn}, J.~C., \& {Johnston}, K.~V.
  2022, \bibinfo{title}{{Similarities behind the high- and
  low-{\ensuremath{\alpha}} disc: small intrinsic abundance scatter and
  migrating stars},} \mnras, 512, 2890, \dodoi{10.1093/mnras/stac610}

\bibitem[{Y.~L. {Lu} {et~al.}(2024){Lu}, {Minchev}, {Buck}, {Khoperskov},
  {Steinmetz}, {Libeskind}, {Cescutti}, {Freeman}, \&
  {Ratcliffe}}]{2024MNRAS.535..392L}
{Lu}, Y.~L., {Minchev}, I., {Buck}, T., {et~al.} 2024, \bibinfo{title}{{There
  is no place like home - finding birth radii of stars in the Milky Way},}
  \mnras, 535, 392, \dodoi{10.1093/mnras/stae2364}

\bibitem[{S.~R. {Majewski} {et~al.}(2017){Majewski}, {Schiavon}, {Frinchaboy},
  {Allende Prieto}, {Barkhouser}, {Bizyaev}, {Blank}, {Brunner}, {Burton},
  {Carrera}, {Chojnowski}, {Cunha}, {Epstein}, {Fitzgerald}, {Garc{\'\i}a
  P{\'e}rez}, {Hearty}, {Henderson}, {Holtzman}, {Johnson}, {Lam}, {Lawler},
  {Maseman}, {M{\'e}sz{\'a}ros}, {Nelson}, {Nguyen}, {Nidever}, {Pinsonneault},
  {Shetrone}, {Smee}, {Smith}, {Stolberg}, {Skrutskie}, {Walker}, {Wilson},
  {Zasowski}, {Anders}, {Basu}, {Beland}, {Blanton}, {Bovy}, {Brownstein},
  {Carlberg}, {Chaplin}, {Chiappini}, {Eisenstein}, {Elsworth}, {Feuillet},
  {Fleming}, {Galbraith-Frew}, {Garc{\'\i}a}, {Garc{\'\i}a-Hern{\'a}ndez},
  {Gillespie}, {Girardi}, {Gunn}, {Hasselquist}, {Hayden}, {Hekker}, {Ivans},
  {Kinemuchi}, {Klaene}, {Mahadevan}, {Mathur}, {Mosser}, {Muna}, {Munn},
  {Nichol}, {O'Connell}, {Parejko}, {Robin}, {Rocha-Pinto}, {Schultheis},
  {Serenelli}, {Shane}, {Silva Aguirre}, {Sobeck}, {Thompson}, {Troup},
  {Weinberg}, \& {Zamora}}]{2017AJ....154...94M}
{Majewski}, S.~R., {Schiavon}, R.~P., {Frinchaboy}, P.~M., {et~al.} 2017,
  \bibinfo{title}{{The Apache Point Observatory Galactic Evolution Experiment
  (APOGEE)},} \aj, 154, 94, \dodoi{10.3847/1538-3881/aa784d}

\bibitem[{C.~R. Malcolm {et~al.}(2026)Malcolm, Grasser, Snellen, de~Regt,
  Picos, Zhang, Stolker, Gandhi, Mollière, Nasedkin, Landman, \&
  Kesseli}]{https://doi.org/10.48550/arxiv.2606.27107}
Malcolm, C.~R., Grasser, N., Snellen, I. A.~G., {et~al.} 2026, The ESO SupJup
  Survey XI. Atmospheric properties of six isolated M- and L-type dwarfs with
  CRIRES+, arXiv, \dodoi{10.48550/ARXIV.2606.27107}

\bibitem[{M. {Mayor} {et~al.}(2011){Mayor}, {Marmier}, {Lovis}, {Udry},
  {S{\'e}gransan}, {Pepe}, {Benz}, {Bertaux}, {Bouchy}, {Dumusque}, {Lo Curto},
  {Mordasini}, {Queloz}, \& {Santos}}]{2011arXiv1109.2497M}
{Mayor}, M., {Marmier}, M., {Lovis}, C., {et~al.} 2011, \bibinfo{title}{{The
  HARPS search for southern extra-solar planets XXXIV. Occurrence, mass
  distribution and orbital properties of super-Earths and Neptune-mass
  planets},} arXiv e-prints, arXiv:1109.2497, \dodoi{10.48550/arXiv.1109.2497}

\bibitem[{J. {Mel{\'e}ndez} {et~al.}(2009){Mel{\'e}ndez}, {Asplund},
  {Gustafsson}, \& {Yong}}]{2009ApJ...704L..66M}
{Mel{\'e}ndez}, J., {Asplund}, M., {Gustafsson}, B., \& {Yong}, D. 2009,
  \bibinfo{title}{{The Peculiar Solar Composition and Its Possible Relation to
  Planet Formation},} \apjl, 704, L66, \dodoi{10.1088/0004-637X/704/1/L66}

\bibitem[{B.~P. {Miller} {et~al.}(2015){Miller}, {Gallo}, {Greene}, {Kelly},
  {Treu}, {Woo}, \& {Baldassare}}]{2015ApJ...799...98M}
{Miller}, B.~P., {Gallo}, E., {Greene}, J.~E., {et~al.} 2015,
  \bibinfo{title}{{X-Ray Constraints on the Local Supermassive Black Hole
  Occupation Fraction},} \apj, 799, 98, \dodoi{10.1088/0004-637X/799/1/98}

\bibitem[{M. {Mittag} {et~al.}(2013){Mittag}, {Schmitt}, \&
  {Schr{\"o}der}}]{2013A&A...549A.117M}
{Mittag}, M., {Schmitt}, J.~H.~M.~M., \& {Schr{\"o}der}, K.-P. 2013,
  \bibinfo{title}{{Ca II H+K fluxes from S-indices of large samples: a reliable
  and consistent conversion based on PHOENIX model atmospheres},} \aap, 549,
  A117, \dodoi{10.1051/0004-6361/201219868}

\bibitem[{J. {Moriarty} {et~al.}(2014){Moriarty}, {Madhusudhan}, \&
  {Fischer}}]{2014ApJ...787...81M}
{Moriarty}, J., {Madhusudhan}, N., \& {Fischer}, D. 2014,
  \bibinfo{title}{{Chemistry in an Evolving Protoplanetary Disk: Effects on
  Terrestrial Planet Composition},} \apj, 787, 81,
  \dodoi{10.1088/0004-637X/787/1/81}

\bibitem[{ {NASA Exoplanet Archive}(2021){NASA Exoplanet Archive}}]{k2pandc}
{NASA Exoplanet Archive}. 2021, K2 Planets and Candidates, Version: 2025-06-29
  15:33 IPAC, \dodoi{10.26133/NEA19}

\bibitem[{J.~E. {Owen} \& A.~P. {Jackson}(2012){Owen} \&
  {Jackson}}]{2012MNRAS.425.2931O}
{Owen}, J.~E., \& {Jackson}, A.~P. 2012, \bibinfo{title}{{Planetary evaporation
  by UV \& X-ray radiation: basic hydrodynamics},} \mnras, 425, 2931,
  \dodoi{10.1111/j.1365-2966.2012.21481.x}

\bibitem[{J.~E. {Owen} \& Y. {Wu}(2013){Owen} \& {Wu}}]{2013ApJ...775..105O}
{Owen}, J.~E., \& {Wu}, Y. 2013, \bibinfo{title}{{Kepler Planets: A Tale of
  Evaporation},} \apj, 775, 105, \dodoi{10.1088/0004-637X/775/2/105}

\bibitem[{I. {Pascucci} {et~al.}(2016){Pascucci}, {Testi}, {Herczeg}, {Long},
  {Manara}, {Hendler}, {Mulders}, {Krijt}, {Ciesla}, {Henning}, {Mohanty},
  {Drabek-Maunder}, {Apai}, {Sz{\H{u}}cs}, {Sacco}, \&
  {Olofsson}}]{2016ApJ...831..125P}
{Pascucci}, I., {Testi}, L., {Herczeg}, G.~J., {et~al.} 2016,
  \bibinfo{title}{{A Steeper than Linear Disk Mass-Stellar Mass Scaling
  Relation},} \apj, 831, 125, \dodoi{10.3847/0004-637X/831/2/125}

\bibitem[{E.~A. {Petigura} {et~al.}(2018){Petigura}, {Marcy}, {Winn}, {Weiss},
  {Fulton}, {Howard}, {Sinukoff}, {Isaacson}, {Morton}, \&
  {Johnson}}]{2018AJ....155...89P}
{Petigura}, E.~A., {Marcy}, G.~W., {Winn}, J.~N., {et~al.} 2018,
  \bibinfo{title}{{The California-Kepler Survey. IV. Metal-rich Stars Host a
  Greater Diversity of Planets},} \aj, 155, 89,
  \dodoi{10.3847/1538-3881/aaa54c}

\bibitem[{M. {Pignatari} {et~al.}(2023){Pignatari}, {Trueman}, {Womack},
  {Gibson}, {C{\^o}t{\'e}}, {Turrini}, {Sneden}, {Mojzsis}, {Stancliffe},
  {Fong}, {Lawson}, {Keegans}, {Pilkington}, {Passy}, {Beers}, \&
  {Lugaro}}]{2023MNRAS.524.6295P}
{Pignatari}, M., {Trueman}, T. C.~L., {Womack}, K.~A., {et~al.} 2023,
  \bibinfo{title}{{The chemical evolution of the solar neighbourhood for
  planet-hosting stars},} \mnras, 524, 6295, \dodoi{10.1093/mnras/stad2167}

\bibitem[{K.~M. {Pontoppidan} {et~al.}(2014){Pontoppidan}, {Salyk}, {Bergin},
  {Brittain}, {Marty}, {Mousis}, \& {{\"O}berg}}]{2014prpl.conf..363P}
{Pontoppidan}, K.~M., {Salyk}, C., {Bergin}, E.~A., {et~al.} 2014,
  \bibinfo{title}{{Volatiles in Protoplanetary Disks},} in Protostars and
  Planets VI, ed. H.~{Beuther}, R.~S. {Klessen}, C.~P. {Dullemond}, \&
  T.~{Henning}, 363--385, \dodoi{10.2458/azu_uapress_9780816531240-ch016}

\bibitem[{K. {Poppenhaeger} \& S.~J. {Wolk}(2014){Poppenhaeger} \&
  {Wolk}}]{2014A&A...565L...1P}
{Poppenhaeger}, K., \& {Wolk}, S.~J. 2014, \bibinfo{title}{{Indications for an
  influence of hot Jupiters on the rotation and activity of their host stars},}
  \aap, 565, L1, \dodoi{10.1051/0004-6361/201423454}

\bibitem[{I. {Ram{\'\i}rez} {et~al.}(2009){Ram{\'\i}rez}, {Mel{\'e}ndez}, \&
  {Asplund}}]{2009A&A...508L..17R}
{Ram{\'\i}rez}, I., {Mel{\'e}ndez}, J., \& {Asplund}, M. 2009,
  \bibinfo{title}{{Accurate abundance patterns of solar twins and analogs. Does
  the anomalous solar chemical composition come from planet formation?},} \aap,
  508, L17, \dodoi{10.1051/0004-6361/200913038}

\bibitem[{J.~G. {Rogers} {et~al.}(2021){Rogers}, {Gupta}, {Owen}, \&
  {Schlichting}}]{2021MNRAS.508.5886R}
{Rogers}, J.~G., {Gupta}, A., {Owen}, J.~E., \& {Schlichting}, H.~E. 2021,
  \bibinfo{title}{{Photoevaporation versus core-powered mass-loss: model
  comparison with the 3D radius gap},} \mnras, 508, 5886,
  \dodoi{10.1093/mnras/stab2897}

\bibitem[{N.~C. {Santos} {et~al.}(2004){Santos}, {Israelian}, \&
  {Mayor}}]{2004A&A...415.1153S}
{Santos}, N.~C., {Israelian}, G., \& {Mayor}, M. 2004,
  \bibinfo{title}{{Spectroscopic [Fe/H] for 98 extra-solar planet-host stars.
  Exploring the probability of planet formation},} \aap, 415, 1153,
  \dodoi{10.1051/0004-6361:20034469}

\bibitem[{K.-T. Shin {et~al.}(2026)Shin, An, Xie, Zhou, \&
  Dai}]{doi:10.1126/science.adu3916}
Shin, K.-T., An, D.-S., Xie, J.-W., Zhou, J.-L., \& Dai, F. 2026,
  \bibinfo{title}{Super-earths and mini-neptunes follow different orbital
  period–eccentricity relations,} Science, 392, 1167,
  \dodoi{10.1126/science.adu3916}

\bibitem[{L. {Su{\'a}rez-Andr{\'e}s} {et~al.}(2017){Su{\'a}rez-Andr{\'e}s},
  {Israelian}, {Gonz{\'a}lez Hern{\'a}ndez}, {Adibekyan}, {Delgado Mena},
  {Santos}, \& {Sousa}}]{2017A&A...599A..96S}
{Su{\'a}rez-Andr{\'e}s}, L., {Israelian}, G., {Gonz{\'a}lez Hern{\'a}ndez},
  J.~I., {et~al.} 2017, \bibinfo{title}{{CNO behaviour in planet-harbouring
  stars. II. Carbon abundances in stars with and without planets using the CH
  band},} \aap, 599, A96, \dodoi{10.1051/0004-6361/201629434}

\bibitem[{T. {Sun} {et~al.}(2023{\natexlab{a}}){Sun}, {Chen}, {Bi}, {Ge},
  {Xiang}, \& {Wu}}]{2023MNRAS.523.1199S}
{Sun}, T., {Chen}, X., {Bi}, S., {et~al.} 2023{\natexlab{a}},
  \bibinfo{title}{{Characterizing abundance-age relations of GALAH stars using
  oxygen-enhanced stellar models},} \mnras, 523, 1199,
  \dodoi{10.1093/mnras/stad1499}

\bibitem[{T. {Sun} {et~al.}(2023{\natexlab{b}}){Sun}, {Ge}, {Chen}, {Bi}, {Li},
  {Zhang}, {Li}, {Wu}, {Bird}, {Ferguson}, {Zhou}, {Ye}, {Long}, \&
  {Zhang}}]{2023ApJS..268...29S}
{Sun}, T., {Ge}, Z., {Chen}, X., {et~al.} 2023{\natexlab{b}},
  \bibinfo{title}{{Age of FGK Dwarfs Observed with LAMOST and GALAH:
  Considering the Oxygen Enhancement},} \apjs, 268, 29,
  \dodoi{10.3847/1538-4365/ace5b0}

\bibitem[{T. {Sun} {et~al.}(2025){Sun}, {Bi}, {Chen}, {Chen}, {Lu}, {Liu},
  {Buck}, {Zhang}, {Li}, {Li}, {Wu}, {Ge}, \& {Ye}}]{2025NatCo..16.1581S}
{Sun}, T., {Bi}, S., {Chen}, X., {et~al.} 2025, \bibinfo{title}{{Potential
  impact of the Sagittarius dwarf galaxy on the formation of young O-rich
  stars},} Nature Communications, 16, 1581, \dodoi{10.1038/s41467-025-56550-1}

\bibitem[{C. {Swastik} {et~al.}(2022){Swastik}, {Banyal}, {Narang}, {Manoj},
  {Sivarani}, {Rajaguru}, {Unni}, \& {Banerjee}}]{2022AJ....164...60S}
{Swastik}, C., {Banyal}, R.~K., {Narang}, M., {et~al.} 2022,
  \bibinfo{title}{{Galactic Chemical Evolution of Exoplanet Hosting Stars: Are
  High-mass Planetary Systems Young?},} \aj, 164, 60,
  \dodoi{10.3847/1538-3881/ac756a}

\bibitem[{C. {Swastik} {et~al.}(2021){Swastik}, {Banyal}, {Narang}, {Manoj},
  {Sivarani}, {Reddy}, \& {Rajaguru}}]{2021AJ....161..114S}
{Swastik}, C., {Banyal}, R.~K., {Narang}, M., {et~al.} 2021,
  \bibinfo{title}{{Host Star Metallicity of Directly Imaged Wide-orbit Planets:
  Implications for Planet Formation},} \aj, 161, 114,
  \dodoi{10.3847/1538-3881/abd802}

\bibitem[{S.~E. Thompson {et~al.}(2018)Thompson, Coughlin, Hoffman, Mullally,
  Christiansen, Burke, Bryson, Batalha, Haas, Catanzarite, Rowe, Barentsen,
  Caldwell, Clarke, Jenkins, Li, Latham, Lissauer, Mathur, Morris, Seader,
  Smith, Klaus, Twicken, Van~Cleve, Wohler, Akeson, Ciardi, Cochran, Henze,
  Howell, Huber, Prša, Ramírez, Morton, Barclay, Campbell, Chaplin,
  Charbonneau, Christensen-Dalsgaard, Dotson, Doyle, Dunham, Dupree, Ford,
  Geary, Girouard, Isaacson, Kjeldsen, Quintana, Ragozzine, Shabram, Shporer,
  Aguirre, Steffen, Still, Tenenbaum, Welsh, Wolfgang, Zamudio, Koch, \&
  Borucki}]{Thompson_2018}
Thompson, S.~E., Coughlin, J.~L., Hoffman, K., {et~al.} 2018,
  \bibinfo{title}{Planetary Candidates Observed by Kepler. VIII. A Fully
  Automated Catalog with Measured Completeness and Reliability Based on Data
  Release 25,} The Astrophysical Journal Supplement Series, 235, 38,
  \dodoi{10.3847/1538-4365/aab4f9}

\bibitem[{P.-W. {Tu} {et~al.}(2025){Tu}, {Xie}, {Chen}, \&
  {Zhou}}]{2025NatAs...9..995T}
{Tu}, P.-W., {Xie}, J.-W., {Chen}, D.-C., \& {Zhou}, J.-L. 2025,
  \bibinfo{title}{{Age dependence of the occurrence and architecture of
  ultra-short-period planet systems},} Nature Astronomy, 9, 995,
  \dodoi{10.1038/s41550-025-02539-1}

\bibitem[{A. {Unni} {et~al.}(2022){Unni}, {Narang}, {Sivarani}, {Manoj},
  {Banyal}, {Surya}, {Rajaguru}, \& {Swastik}}]{2022AJ....164..181U}
{Unni}, A., {Narang}, M., {Sivarani}, T., {et~al.} 2022,
  \bibinfo{title}{{Carbon Abundance of Stars in the LAMOST-Kepler Field},} \aj,
  164, 181, \dodoi{10.3847/1538-3881/ac8b7c}

\bibitem[{F. {Wanderley} {et~al.}(2025){Wanderley}, {Cunha}, {Smith}, {Souto},
  {Pascucci}, {Behmard}, {Allende Prieto}, {Beaton}, {Bizyaev}, {Daflon},
  {Hasselquist}, {Howell}, {Majewski}, \& {Pinsonneault}}]{2025ApJ...993..233W}
{Wanderley}, F., {Cunha}, K., {Smith}, V.~V., {et~al.} 2025,
  \bibinfo{title}{{An Analysis of the Radius Gap in a Sample of Kepler, K2, and
  TESS Exoplanets Orbiting M-dwarf Stars},} \apj, 993, 233,
  \dodoi{10.3847/1538-4357/ae058e}

\bibitem[{T. Wang {et~al.}(2025)Wang, Yuan, Chen, Xiang, Zhang, Huang, Gu,
  Wang, \& Li}]{Wang_2025}
Wang, T., Yuan, H., Chen, B., {et~al.} 2025, \bibinfo{title}{An All-sky 3D Dust
  Map Based on Gaia and LAMOST,} The Astrophysical Journal Supplement Series,
  280, 15, \dodoi{10.3847/1538-4365/adea39}

\bibitem[{M. {Xiang} \& H.-W. {Rix}(2022){Xiang} \&
  {Rix}}]{2022Natur.603..599X}
{Xiang}, M., \& {Rix}, H.-W. 2022, \bibinfo{title}{{A time-resolved picture of
  our Milky Way's early formation history},} \nat, 603, 599,
  \dodoi{10.1038/s41586-022-04496-5}

\bibitem[{M. {Xiang} {et~al.}(2019){Xiang}, {Ting}, {Rix}, {Sandford}, {Buder},
  {Lind}, {Liu}, {Shi}, \& {Zhang}}]{2019ApJS..245...34X}
{Xiang}, M., {Ting}, Y.-S., {Rix}, H.-W., {et~al.} 2019,
  \bibinfo{title}{{Abundance Estimates for 16 Elements in 6 Million Stars from
  LAMOST DR5 Low-Resolution Spectra},} \apjs, 245, 34,
  \dodoi{10.3847/1538-4365/ab5364}

\bibitem[{J. {Yang} {et~al.}(2026){Yang}, {Kempton}, \&
  {Savel}}]{2026ApJ..1003L..16Y}
{Yang}, J., {Kempton}, E. M.-R., \& {Savel}, A.~B. 2026,
  \bibinfo{title}{{Sub-Neptunes as Soot Factories: Deep Atmosphere Hydrocarbon
  Formation and Quenching as the Origin of Sub-Neptune Aerosol Trends},} \apjl,
  1003, L16, \dodoi{10.3847/2041-8213/ae6914}

\bibitem[{J.-Y. {Yang} {et~al.}(2023){Yang}, {Chen}, {Xie}, {Zhou}, {Dong},
  {Zhu}, {Zheng}, {Liu}, {Zong}, \& {Luo}}]{2023AJ....166..243Y}
{Yang}, J.-Y., {Chen}, D.-C., {Xie}, J.-W., {et~al.} 2023,
  \bibinfo{title}{{Planets Across Space and Time (PAST). IV. The Occurrence and
  Architecture of Kepler Planetary Systems as a Function of Kinematic Age
  Revealed by the LAMOST-Gaia-Kepler Sample},} \aj, 166, 243,
  \dodoi{10.3847/1538-3881/ad0368}

\bibitem[{M. {Zhang} {et~al.}(2025){Zhang}, {Xiang}, {Ting}, {Amarsi}, {Zhang},
  {Shi}, {Yuan}, {Li}, {Wang}, {Wu}, {Wu}, {Mou}, {Yan}, \&
  {Liu}}]{2025ApJS..279....5Z}
{Zhang}, M., {Xiang}, M., {Ting}, Y.-S., {et~al.} 2025,
  \bibinfo{title}{{Homogeneous Stellar Atmospheric Parameters and 22 Elemental
  Abundances for FGK Stars Derived from LAMOST Low-resolution Spectra with
  DD-PAYNE},} \apjs, 279, 5, \dodoi{10.3847/1538-4365/add016}

\bibitem[{W. Zhu \& S. Dong(2021)Zhu \& Dong}]{Zhu2021}
Zhu, W., \& Dong, S. 2021, \bibinfo{title}{Exoplanet Statistics and Theoretical
  Implications,} Annual Review of Astronomy and Astrophysics, 59, 291–336,
  \dodoi{10.1146/annurev-astro-112420-020055}

\end{thebibliography}
\bibliographystyle{aasjournalv7}

\end{document}